\documentclass[usenatbib,useAMS]{mnras}

\usepackage{newtxtext,newtxmath}

\usepackage[T1]{fontenc}
\usepackage{ae,aecompl}
\usepackage[dvipdfmx]{graphicx}	
\usepackage{amsmath}	
\usepackage{amsfonts}
\usepackage{multicol}        
\usepackage{bm}		
\usepackage{pdflscape}	
\usepackage{epstopdf}
\usepackage{comment}
\usepackage{xcolor}

\usepackage{fancyvrb}
\usepackage{fancybox}

\newcommand{\mtrv}[1]{{\textcolor{black}{#1}}}
\newcommand{\tkrv}[1]{{\textcolor{black}{#1}}}
\newcommand{\tkrrv}[1]{{\textcolor{black}{#1}}}

\newcommand{\beq}{\begin{equation}}
\newcommand{\beqa}{\begin{eqnarray}}
\newcommand{\eeq}{\end{equation}}
\newcommand{\eeqa}{\end{eqnarray}}

\newcommand{\bx}{\bm{x}}

\newcommand{\bs}{\bm{s}}
\newcommand{\bk}{\bm{k}}

\newcommand{\bn}{\bm{n}}

\newcommand{\hn}{\hat{n}}

\newcommand{\hMpci}{$h\,$Mpc$^{-1}$}

\usepackage{siunitx}

\def\avrg#1{\left\langle #1 \right\rangle}

\title[Power spectrum of halo intrinsic alignments in simulations]{Power spectrum of halo intrinsic alignments in simulations}

\author[Kurita et al.]
{Toshiki~Kurita$^{1,2}$\thanks{E-mail: toshiki.kurita@ipmu.jp}, 
Masahiro~Takada$^{1}$,
Takahiro~Nishimichi$^{3,1}$,
Ryuichi~Takahashi$^{4}$,
\newauthor 
Ken~Osato$^5$,
Yosuke~Kobayashi$^{1,2}$
\\
$^1$ Kavli Institute for the Physics and Mathematics of the Universe (WPI),
The University of Tokyo Institutes for Advanced Study (UTIAS),
\\ The University of Tokyo, 5-1-5 Kashiwanoha, Kashiwa-shi, Chiba, 277-8583, Japan\\
$^{2}$ Department of Physics, The University of Tokyo, 7-3-1 Hongo, Bunkyo-ku, Tokyo 113-0033 Japan\\
$^{3}$ Center for Gravitational Physics, Yukawa Institute for Theoretical Physics, Kyoto University, Kyoto 606-8502, Japan\\
${}^4$ Faculty of Science and Technology, Hirosaki University, 3 Bunkyo-cho, Hirosaki, Aomori 036-8561, Japan\\
${}^5$ Institut d'Astrophysique de Paris, Sorbonne Universit\'{e}, CNRS, UMR 7095, 75014 Paris, France
}

\begin{document}
\label{firstpage}
\pagerange{\pageref{firstpage}--\pageref{lastpage}}

\thisfancyput(12.5cm,0.5cm){\large{IPMU20-0044, YITP-20-48}}

\maketitle

\begin{abstract}
We use a suite of $N$-body simulations to study intrinsic alignments (IA) of halo shapes with the surrounding large-scale structure in the $\Lambda$CDM model.
For this purpose, we develop a novel method to measure multipole moments of the three-dimensional power spectrum of the $E$-mode field of halo shapes with the matter/halo distribution, $P_{\delta E}^{(\ell)}(k)$ (or $P^{(\ell)}_{{\rm h}E}$), and those of the auto-power spectrum of the $E$ mode, $P^{(\ell)}_{EE}(k)$, based on the $E$/$B$-mode decomposition. 
The IA power spectra have non-vanishing amplitudes over the linear to nonlinear scales, and the large-scale amplitudes at $k\lesssim 0.1~h~{\rm Mpc}^{-1}$ are related to the matter power spectrum via a constant coefficient ($A_{\rm IA}$), similar to the linear bias parameter
\mtrv{of galaxy or halo density field.} 
We find that the cross- and auto-power spectra $P_{\delta E}$ and $P_{EE}$ at nonlinear scales, $k\gtrsim 0.1~h~{\rm Mpc}^{-1}$, show different $k$-dependences relative to the matter power spectrum, suggesting a violation of the nonlinear alignment model commonly used to model contaminations of cosmic shear signals.  
The IA power spectra exhibit baryon acoustic oscillations, and vary with halo samples of different masses, redshifts and cosmological parameters ($\Omega_{\rm m}, S_8$). 
The cumulative signal-to-noise ratio for the IA power spectra is about 60\% of that for the halo density power spectrum, where the super-sample covariance is found to give a significant contribution to the total covariance.
Thus our results demonstrate that the IA power spectra of galaxy shapes, measured from imaging and spectroscopic surveys for an overlapping area of the sky, can be \mtrv{used}
to probe the underlying matter power spectrum, the primordial curvature perturbations, and cosmological parameters, in addition to the standard galaxy density power spectrum. 
\end{abstract}




\section{Introduction}
\label{sec:intro}

There are many ongoing and planned imaging and spectroscopic surveys covering a wide area of the sky \citep[e.g.,][]{2014PASJ...66R...1T}. 
These surveys aim to address the fundamental questions in cosmology: properties of the primordial perturbations that are seeds of the present-day cosmic structures, and the physical nature of dark matter and dark energy that are introduced to explain the dominant source of gravity and the cosmic accelerating expansion in the late-time universe \citep[e.g., see][for a review]{2013PhR...530...87W}.

Cosmological observables for spectroscopic galaxy surveys, which have been extensively studied in the literature, are the clustering statistics of galaxy distribution in angular or redshift space \citep{2005ApJ...633..560E,2017MNRAS.470.2617A}. 
Those for imaging surveys are weak lensing distortion in galaxy images, the so-called cosmic shear, which probes the matter distribution in foreground large-scale structures \citep{2017MNRAS.465.1454H,2018PhRvD..98d3528T,2019PASJ...71...43H,2020PASJ...72...16H}. 
The joint probes combining the galaxy clustering and the weak lensing are proven as a powerful means of constraining cosmological parameters, by breaking parameter degeneracies, especially circumventing the impact of galaxy bias uncertainty on cosmological constraints \citep{2015ApJ...806....2M,2018PhRvD..98d3526A}.

The cold dark matter (CDM) dominated structure formation model predicts that shapes of galaxies interact with the surrounding gravitational (tidal) field in large-scale structure, and it induces intrinsic (not lensing-induced) correlations between galaxy shapes in the common large-scale structure, so-called intrinsic alignments (IA) \citep{2000ApJ...545..561C,Catelanetal2001,2002ApJ...568...20C,2002MNRAS.335L..89J}. 
Usually the IA effect is considered as one of the most important physical systematic effects in the cosmic shear analysis \citep{HirataSeljak2004,2006MNRAS.371..750H} \citep[see also][for a review]{Joachimietal2015,Kiesslingetal2015,Kirketal2015,TroxelandIshak2015}. 
Only very recently several theoretical works have started considering the IA effects as cosmological signals \citep{2012PhRvD..86h3513S,ChisariDvorkin2013,Schmidtetal2015,2018JCAP...08..014K,2019PhRvD.100j3507O,2020MNRAS.494..694O,2020MNRAS.493L.124O,2020arXiv200105962T}.
The IA correlations have been indeed measured from observational data, especially for early-type red galaxies \citep{Mandelbaumetal2006,Okumuraetal2009,Singhetal2015,2019A&A...624A..30J,2019MNRAS.489.5453S,2020arXiv200209826Y}.

Based on the above background, there have been analytical and numerical attempts to develop an accurate model \mtrv{or advance the physical understanding} of the IA effects. 
For an analytical approach it is usually assumed that galaxy shapes are tracers of the underlying gravitational tidal field that is sourced by the total matter (mainly dark matter) distribution in large-scale structure, and this model is called the linear (tidal) alignment model \citep{HirataSeljak2004}. 
Then, the linear theory or perturbation theory of structure formation can be used to express the IA correlations in terms of the power spectrum or higher-order moments of matter and tidal fields \citep[also see][]{Blazeketal2015,Schmitzetal2018,Blazeketal2017,2020JCAP...01..025V}. 
For the cosmic shear analysis, an empirical model, the so-called nonlinear alignment model \citep{2007NJPh....9..444B}, is often used to model the IA contamination to the cosmic shear signals at nonlinear scales, where the linear matter power spectrum appearing in the IA correlation is replaced with the nonlinear matter power spectrum. 
There are also simulation-based studies using cosmological $N$-body simulations \citep{2002MNRAS.335L..89J,2017ApJ...848...22X,2018MNRAS.474.1165P,Osatoetal2018,2017arXiv170608860O,2019PhRvD.100j3507O,2020MNRAS.494..694O,2020arXiv200203867S} and cosmological hydrodynamical simulations \citep{2015MNRAS.448.3522T,2015MNRAS.452.3369C,2015MNRAS.453..721V,2015MNRAS.454.2736C,2015MNRAS.454.3328V,2017MNRAS.472.1163C,2018MNRAS.476.3460T,2020arXiv200900276S}. 
Moreover, the halo model approach has been recently developed to model the IA effects of galaxies at nonlinear scales, more specifically inside the host halos \citep{2010MNRAS.402.2127S,2020arXiv200302700F}.

However, most of the previous studies are on the real- or configuration-space IA correlations, except for the perturbation theory based studies \citep[e.g.,][]{Blazeketal2017}. 
Hence the purpose of this paper is to develop a novel {\it method} to measure the {\it three-dimensional} power spectrum of the IA effects, using the $E$/$B$-mode decomposition method developed in the cosmic microwave polarization and the cosmic shear. 
We then apply the method to shapes of halos measured from a suite of $N$-body simulations, generated in \citet{2019ApJ...884...29N}, and estimate the auto-power spectra of the halo shape $E$/$B$ modes and the cross-power spectrum of the $E$ mode with the surrounding matter or halo distribution. 
Since the halo shapes are a spin-2 field defined in the two-dimensional plane perpendicular to the line-of-sight direction, the IA power spectra break the statistical isotropy and display anisotropic modulations depending on the angle between wavevector and the line-of-sight direction, just like the redshift-space power spectrum of galaxies. 
We use the measured IA power spectra to study a validity of the linear and nonlinear alignment models, the baryon acoustic oscillations, the information content (the cumulative signal-to-noise ratio) and the redshift-space distortion effect, compared to the standard power spectrum of halo density field. 
We also examine how the IA power spectra vary with halo samples of different masses, redshifts and cosmological parameters. 
In doing these, we pay special attention to the fact that keeping the three-dimensional Fourier modes in the IA power spectrum measurements enables one to extract the full information of \mtrv{IA effects at the two-point statistics level}, compared to the angular or projected correlation functions that are often studied in analogy to the cosmic shear correlations. 
The method developed in this paper can be applied to imaging and spectroscopic galaxy surveys observing the same area of the sky, where galaxy shapes are measured from the imaging data and the three-dimensional positions of galaxies are obtained from the spectroscopic data. 
This is the case, e.g., for the BOSS survey combined with the Subaru HSC survey \citep{2018PASJ...70S...4A}, the Subaru HSC/PFS surveys \citep{2014PASJ...66R...1T}, the ESA \textit{Euclid}\footnote{\url{https://www.cosmos.esa.int/web/euclid}} and 
\mtrv{the NASA Roman Space Telescope\footnote{\url{https://roman.gsfc.nasa.gov}}.}
\mtrv{We would like to again stress the standing point of this paper; we develop the method of measuring the IA power spectrum, and then study 
how the IA effect can be used as a cosmological {\it signal}. In other words, we will not consider the IA effect as a contaminating effect on the cosmic shear that is the main focus of most previous studies.}

This paper is structured as follows. 
In Section~\ref{sec:ia-model}, we review the intrinsic alignment model, mainly the tidal/linear alignment model, and define notations and quantities used in this paper.
In Section~\ref{sec:estimate-shear-field}, we give details of our simulations and describe the methods to measure the ellipticities of dark matter halos and the IA power spectra from the ellipticity/shear field. 
In Section~\ref{sec:results}, we present our results.
We give conclusion and discussion in Section~\ref{sec:conclusion}.
 

\section{Intrinsic alignment model}
\label{sec:ia-model}

\subsection{Preliminaries}
\label{subsec:preiminaries}

Here we briefly review the IA model in large-scale structure. 
The IA model is based on the assumption that the shear tensor, defined by shapes of galaxies or halos at a redshift $z$, $g_{ij}(\bx; z)$, originates from the gravitational tidal tensor at a redshift $z_{\rm IA}$ higher than $z$ around the epoch of the formation of the galaxy of interest, i.e., 
\begin{equation}
	g_{ij}(\bm{x}; z) \propto K_{ij}(\bm{x}; z_{\rm IA}),
	\label{eq:def_IA_relation}
\end{equation}
where
\begin{equation}
	K_{ij}(\bm{x}; z) \equiv \tkrv{\frac{1}{4\pi G\bar{\rho}_{\rm m}(z)a^2}}\left( \nabla_i \nabla_j - \frac{1}{3}\delta_{ij} \nabla^{2} \right) \Phi(\bm{x}; z),
	\label{eq:Kij_def}
\end{equation}
and 
\tkrv{$a$ is the scale factor ($a=1/(1+z)$), $\bar{\rho}_{\rm m}(z)$ is the mean mass density at redshift $z$, 
}
$\Phi(\bm{x},z)$ is the gravitational potential field, or the metric perturbation in the General Relativity framework. 
As stressed in \citet{HirataSeljak2004}, the relationship of Eq.~(\ref{eq:def_IA_relation}) is expected to hold only on large scales in the linear regime, in analogy with the linear bias model that relates the spatial distributions of galaxies and matter on large scales via a proportionality factor, i.e., a linear bias coefficient.
The gravitational potential field is related to the mass density fluctuation field via the Poisson equation as
\begin{equation}
\nabla^2 \Phi(\bm{x};z)=4\pi G\bar{\rho}_{\rm m}(z) a^2\delta(\bm{x};z).
\label{eq:Poisson}
\end{equation}
where 
$\delta(\bm{x};z)$ is the mass density fluctuation field.
\mtrv{Note that the tidal field $K_{ij}$, defined by the above equation, has the same dimension as the mass density fluctuation, $\delta(\bm{x})$.}

We can observe the ``shape'' of individual galaxies projected onto the sky, which is on a two-dimensional plane perpendicular to the line-of-sight direction, under the flat-sky approximation (this would be a good approximation as a galaxy size is very small compared to the curvature scale of the celestial sphere). 
In other words we cannot observe a three-dimensional shape of the galaxy. 
Hence we define an ``observed'' shear of a galaxy or halo as
\begin{equation}
    \gamma_{ij}(\bm{x}; z) \equiv \left(\mathcal{P}_{ik}(\bm{\hat{n}}) \mathcal{P}_{jl}(\bm{\hat{n}}) - \frac{1}{2} \mathcal{P}_{ij}(\bm{\hat{n}}) \mathcal{P}_{kl}(\bm{\hat{n}}) \right) g_{kl}(\bm{x}; z),
\end{equation}
where $\hat{\bm{n}}$ is the unit vector of the line-of-sight direction, and $\mathcal{P}_{ij}(\bm{\hat{n}}) \equiv \delta_{ij} - \hat{n}_i \hat{n}_j$ that is the projection tensor onto the plane perpendicular to the line-of-sight direction.
Throughout this paper, we refer to the coordinate components as $\bm{x}=(x^1,x^2,x^3)$ and do not use $x^3=z$ to avoid confusion with redshift ``$z$''. 
If we set the $x^3$-direction to the line-of-sight direction, i.e., $\hat{\bm{n}} \parallel x^3$, $\gamma_{ij}$ is expressed as
\begin{equation}
	\gamma_{ij}(\bm{x}; z) = 
	\begin{pmatrix}
		\gamma_+ & \gamma_{\times} & 0\\
		\gamma_{\times} & -\gamma_+ & 0\\
		0 & 0 & 0
	\end{pmatrix} . 
\end{equation}
Since the shear tensor is traceless and symmetric, $\gamma_{ij}$ has two degrees of freedom for which we introduce the two components, $\gamma_{+,\times}$, in analogy to the weak lensing shear \citep{BartelmannSchneider2001,Dodelson2017}.
The IA model relates the shear tensor to the tidal field as
\begin{align}
	\gamma_+(\bm{x}; z) &\equiv -\frac{C_1}{4\pi G}(\nabla_1^2-\nabla_2^2)\Phi(\bm{x}; z_{\rm IA}), \nonumber\\
	\gamma_\times(\bm{x}; z) & \equiv -\frac{2C_1}{4\pi G}\nabla_1\nabla_2 \Phi({\bm{x}; z_{\rm IA}}).
	\label{eq:la}
\end{align}
Throughout this paper we adopt the plane-parallel or distant-observer approximation.
Following the convention in the literature \citep{HirataSeljak2004}, we introduced a prefactor $C_1/4\pi G$, and $C_1$ is a constant factor that has a dimension of $\rho^{-1}$.
$C_1$ is a proportionality factor that depends on properties of sample galaxies or halos, e.g., luminosity (for galaxies), mass, redshift, cosmological parameters and so on. 
The minus sign is conventionally taken so that, if shapes of galaxies and halos are elongated along the direction of the mass accretion from the surrounding structures,  $C_1$ turns to be positive.

Using the Poisson equation, Eq.~(\ref{eq:la}) can be expressed in Fourier space as
\begin{equation}
	\gamma_{(+,\times)}(\bm{k},z) = -C_1 \Omega_{\rm m}\rho_{\rm cr0} (1+z_{\rm IA}) f_{(+,\times)}(\bm{\hat{k}}) \delta(\bm{k}, z_{\rm IA}),
	\label{eq:laF}
\end{equation}
where $\rho_{\rm cr0}$ is the critical density today and we have defined the function $f_{(+,\times)}$, following \cite{Blazeketal2015}, as
\begin{equation}
	f_{(+,\times)}(\bm{\hat{k}}) \equiv  (1-\mu^2)(\cos 2\phi, \sin 2\phi), 
\end{equation}
where $\mu \equiv \bm{\hat{n}} \cdot \bm{\hat{k}} = \hat{k}_3$ and $\phi = \tan^{-1}({\hat{k}_1/\hat{k}_2})$. 
The factor $(1-\mu^2)$ in the kernel, $f_{(+,\times)}$, reflects the fact that the IA shear arises from Fourier modes in two-dimensional plane perpendicular to the line-of-sight direction, $\bk_\perp$. 
For example, Fourier modes along the line-of-sight direction, which have $\mu=\pm 1$, do not cause the observed IA shear. 
This is opposite to the redshift-space distortion (RSD)
due to peculiar velocities of galaxies, which arise from Fourier modes along the line-of-sight direction. 
If we take $z_{\rm IA}$ to be in the matter dominated era for an epoch of the IA generation, the amplitude of tidal field on linear scales is constant in time, and therefore the IA reflects the primordial tidal field. 
This model is called the {\it primordial alignment} model, and in this case we have
\begin{equation}
	\gamma_{(+,\times)}(\bm{k},z) = -A_{\rm IA}C_1 \rho_{\rm cr0} \frac{\Omega_{\rm m}}{D(z)} f_{(+,\times)}(\bm{\hat{k}}) \delta(\bm{k}, z),
	\label{eq:def_of_AI}
\end{equation}
where $D(z)$ is the linear growth factor. 
We set $C_1 \rho_{\rm cr0}=0.0134$ and we employ the normalization $D(z=0)=1$ in this work, following the convention in \cite{Joachimi2011}. 
We use the dimensionless parameter $A_{\rm IA}$ to characterize the amplitude of the IA signal.

For an practical measurement of the IA effect, we further take into account the RSD effect 
caused by peculiar velocities of galaxies or halos. 
We will discuss the RSD effect in a separate section later.

\subsection{$E$/$B$ decomposition of the IA power spectrum}
\label{subsec:eb_decomposition}

As we described, the galaxy shape induced by the IA effect is a spin-2 field by definition. 
This is a useful property, and we can use the $E$/$B$-mode decomposition of the observed galaxy shape field that gives a unique decomposition of the two degrees of freedom in the spin-2 field. 
The $E$ mode is a physical mode caused by the scalar gravitational potential, and the $B$ mode is a non-physical mode that cannot be generated by the scalar mode in the linear regime, so served as an indicator of systematic errors in actual measurements. 
However, note that the $E$ and $B$ modes are mixed in the nonlinear regime or if the field is a nonlinear field of the underlying scalar fields, which indeed occurs in the IA power spectrum as we will show later.
In analogy with CMB polarization \citep{1997PhRvD..55.1830Z,1997PhRvD..55.7368K} and weak lensing \citep{2002ApJ...568...20C}, the $E$/$B$-mode decomposition is non-local in real space, while it is ``local'' in Fourier space. 
From Eq.~(\ref{eq:laF}), we can define the $E$/$B$ modes of galaxy shapes, denoted as $\gamma_{E, B}$:
\begin{align}
	\gamma_E(\bm{k}) &=  \gamma_{+}(\bm{k}) \cos{2\phi}	+  \gamma_{\times}(\bm{k}) \sin{2\phi},
	\label{eq:gammaE}\\
	\gamma_B(\bm{k}) &= - \gamma_{+}(\bm{k}) \sin{2\phi} +  \gamma_{\times}(\bm{k}) \cos{2\phi}.
	\label{eq:gammaB}
\end{align}

From these equations, in this paper we consider the following 3D power spectra to study the IA effect:
\begin{align}
    \avrg{\gamma_E(\bm{k})\delta(\bm{k}')}&\equiv P_{\delta E}(\bm{k})(2\pi)^3\delta^3_D(\bm{k}+\bm{k}'),\nonumber\\
    \avrg{\gamma_E(\bm{k})\gamma_E(\bm{k}')}&\equiv P_{EE}(\bm{k})(2\pi)\delta^3_D(\bm{k}+\bm{k}'),
    \label{eq:def_PdE} 
\end{align}
where $\delta^3_D(\bm{k})$ is the 3D Dirac delta function, $P_{\delta E}(\bm{k})$ is the cross-power spectrum between the mass density field and the $E$ mode of galaxy shape, and $P_{EE}(\bm{k})$ is the auto-power spectrum of the $E$-mode field. 
We should emphasize that, although the $E$/$B$ modes are defined with respect to the Fourier modes $\bm{k}_\perp$ in the ``two''-dimensional plane perpendicular to the line-of-sight direction, the power spectra are given as a function of the three-dimensional wavevector, $\bm{k}$. 
In addition, the power spectra are not only a function of the scalar $k(=|\bm{k}|)$, but also depends on the direction of $\bm{k}$. 
These 3D power spectra contains the full information on the IA effect at the level of two-point statistics. 
\mtrv{In a conventional method that has been used in the actual measurement, the projected correlation function is used to measure the IA effect, where 
the correlation function is obtained by}
integrating the above 3D power spectrum information along the line-of-sight direction, in analogy with the weak lensing measurement. 
As we will show, this projection leads to a loss of the underlying information. 
For the $B$-mode power spectra, $\avrg{\gamma_B\gamma_B}=0$ for the IA caused by the scalar tidal field in the linear regime, and $\avrg{\gamma_B\delta}=\avrg{\gamma_E\gamma_B}=0$ due to the statistical parity invariance. 
These give a useful sanity check of residual systematic errors in actual measurements.

For convenience of our discussion, we define the multipole moments of the IA power spectrum as
\begin{equation}
	P^{(\ell)}_{XY}(k) \equiv \frac{2\ell +1}{2}\int^1_{-1} {\rm d}\mu \mathcal{L}_{\ell}(\mu)P_{XY}(k,\mu),
\end{equation}
where the subscripts $X$ and $Y$ are labels for $\delta$ (density), $\mathrm{h}$ (halos), or $E$ (shape), and $\mathcal{L}_{\ell}(x)$ is the Legendre polynomial of order $\ell$.
Due to the geometrical nature of $E$/$B$ modes of the intrinsic galaxy shapes, the following relations between the multipole moments are expected to hold \citep[see][for those in configuration space]{2020MNRAS.493L.124O}: 
\begin{equation}
	\frac{P^{(2)}_{\delta E}}{P^{(0)}_{\delta E}} = -1, \frac{P^{(2)}_{EE}}{P^{(0)}_{EE}} = -\frac{10}{7},~\frac{P^{(4)}_{EE}}{P^{(0)}_{EE}} = \frac{3}{7}.
	\label{eq:relations_moments}
\end{equation}
For the cross-power spectrum, $P_{\delta E}$ or $P_{{\rm h}E}$, the above relation always holds because it comes from the geometrical factor $(1-\mu^2)$ in the definition of the \textit{projected} shapes, and thus does not rely on the specific IA model
\citep[also see][]{2020MNRAS.493L.124O}.
On the other hand, for the auto-power spectrum, $P_{EE}$, the above relation holds in the linear regime to a good approximation, but is not exact even in the linear regime (small $k$) due to the nonlinear shot-noise contribution (see below). 
Note that the projection effects do not cause the higher-order moments beyond the 2nd- or 4th-order moments for $P_{\delta E}(P_{{\rm h}E})$ and $P_{EE}$ in real space, respectively. 

Plugging Eq.~(\ref{eq:laF}) into Eq.~(\ref{eq:def_PdE}), we find that the linear IA model predicts the power spectra to be given as
\begin{align}
	P_{\delta E}(k, \mu; z) &= -A_{\rm IA}C_1 \rho_{\rm cr0} \frac{\Omega_{\rm m}}{D(z)} (1-\mu^2) P^{\rm lin}_{\delta}(k; z),\label{eq:PdE}\\
	P_{EE}(k, \mu; z) &= \left[ A_{\rm IA}C_1 \rho_{\rm cr0} \frac{\Omega_{\rm m}}{D(z)} (1-\mu^2) \right]^2 P^{\rm lin}_{\delta}(k; z),\label{eq:PEE}
\end{align}
where $P^{\rm lin}_{\delta}(k; z)$ is the linear matter power spectrum. 
This is called as the {\it linear} alignment model. 
If we replace $P^{\rm lin}_\delta(k)$ with the nonlinear matter power spectrum, denoted as $P^{\rm NL}_\delta(k)$, it gives the {\it nonlinear} alignment model \citep{2007NJPh....9..444B}, which has been often used in the weak lensing cosmology analysis \citep[e.g.,][]{2019PASJ...71...43H}.
These alignment models predict the specific relations between $P_{\delta E}$ and $P_{EE}$ via the same coefficient with respect to the matter power spectrum.
The above equations are found to satisfy Eq.~(\ref{eq:relations_moments}). 

Note that the shear field estimated by using shapes of galaxies or halos $\tilde{\gamma}_{ij}$ is a density-weighted field because we can sample the shape field only at the positions of halos/galaxies and the halos/galaxies are biased tracers of the underlying matter density field, i.e., $\tilde{\gamma}_{ij} = (1+\delta_{\rm g/h}) \gamma_{ij}$ (see Appendix~\ref{app:density_weighted_field} for details) \citep[also see][for a similar discussion on the redshift-space distribution field of galaxies]{2011JCAP...11..039S}. 
At the leading order, its Fourier transform is written as
\begin{align}
	\tilde{\gamma}_{ij}(\bk) &= \left[(1 + b_{\rm g/h}\delta^{\rm lin}) \ast \gamma_{ij}\right]\!(\bk)~\nonumber\\
	&= \int\!\!\frac{{\rm d}^3\bk'}{(2\pi)^3} \gamma_{ij}(\bk-\bk')\left[(2\pi)^3\delta_D(\bk') + b_{\rm g/h}\delta^{\rm lin}(\bk')\right],
	\label{eq:gamma_weighted}
\end{align}
where $b_{g/h}$ is a linear galaxy/halo bias. 
Therefore $P_{EE}$ and $P_{BB}$ have $\mathcal{O}((P_\delta^{\rm lin})^2)$ correction terms in addition to Eq.~(\ref{eq:PEE}).
These nonlinear terms of the fluctuation fields lead to a leakage of $E$ mode into $B$ mode, as we will discuss below.

In order to predict the IA effect beyond linear theory, one might want to use the perturbation theory of structure formation \citep{2002PhR...367....1B} or the effective field theory of large-scale structure \citep{2009JCAP...08..020M,2012JCAP...07..051B}. 
For this kind of approach, one can write down a general expansion of the IA field in terms of series of the underlying matter fields and possibly additional counter terms, with coefficients for each term \citep[see][for recent works]{Blazeketal2015,Schmidtetal2015,Blazeketal2017,Schmitzetal2018,2020JCAP...01..025V}.

\section{Measurement method of IA power spectra from $N$-body simulations}
\label{sec:estimate-shear-field}

In this section, we describe details of $N$-body simulations and the halo catalogs, the way to quantify shapes of halos, and the way to measure the IA power spectra from the simulations.

\subsection{$N$-body simulations and halo catalogs}
\label{subsec:n-body}

In this paper, we use a subset of the $N$-body simulation data in \texttt{Dark Quest} \citep{2019ApJ...884...29N}, more exactly the high-resolution (\texttt{HR}) suite constructed in the paper, and the associated halo catalogs. 
\tkrrv{
We generate the initial conditions using {\tt CAMB} \citep{camb} to compute the linear matter power spectrum at the initial redshift and {\tt 2LPTIC} \citep{scoccimarro98,crocce06a,crocce06b,nishimichi09,Valageas11a,valageas11b} to set up the initial displacement and velocity of each $N$-body particle and then simulate the particle distribution using {\tt Gadget2} \citep[][]{gadget2} with $2048^3$ particles in a periodic cubic box size of $1~h^{-1}~{\rm Gpc}$.
We employ the flat-$\Lambda$CDM model with the following values of cosmological parameters for the fiducial cosmology:
$(\omega_{\rm b},\omega_{\rm c},\Omega_{\rm \Lambda},\ln(10^{10}A_{\rm s}),n_{\rm s})=(0.02225,0.1198,0.6844,3.094,0.9645)$, which are consistent with the {\it Planck} results \citep{planck-collaboration:2015fj}.
The mass of $N$-body particle corresponds to $1.02\times 10^{10}~h^{-1}M_\odot$ for the fiducial cosmology.}

For each simulation realization, we identify halos in the post-processing computation, using a phase space finder, {\tt Rockstar} \citep{Behroozi:2013}. 
The center of each halo is estimated from the center-of-mass location of a subset of member particles in the inner part of halo, which is considered as a proxy of the mass density maximum.
\tkrv{Throughout this paper, we use
the virial mass in the {\tt Rockstar} outputs as
the mass of each halo; $M_{\rm h} \equiv M_{\rm vir}$. We use halos with masses greater than $10^{12}~h^{-1}M_\odot$, and}
use the outputs of $N$-body realizations at 21 redshifts in the range of $0\le z\le 1.48$, evenly stepped by the linear growth 
factor for the fiducial {\it Planck} cosmology \citep[see][for details]{2019ApJ...884...29N}.

\subsection{Measurements of halo shapes}
\label{subsec:shape_measurement}

We now need to quantify the ``shape'' of individual halos. Since dark matter halos are not relaxed nor in dynamical equilibrium and do not have any clear boundary, there is no unique definition of halo shape. 
What we can observe from data is only 
the ``shape'' of a galaxy, or that of \tkrv{stellar} distribution, and those stars would form in the center around the mass density maximum in each host halo due to baryonic dissipative processes forming stars.
Hence, in order to estimate a ``central-galaxy-like'' shape of each halo, we use the following inertia tensor of $N$-body particle distribution in each halo \citep{Osatoetal2018} \citep[also see][for the similar definition]{2012MNRAS.420.3303B,2015MNRAS.448.3522T}: 
\begin{equation}
	I_{ij} \tkrv{\propto} \sum_p w_p\Delta x^i_p \Delta x^j_p,
	\label{eq:Iij}
\end{equation}
where $\Delta \bx_p \equiv \bx_p - \bx_{\rm h}$, $\bm{x}_{\rm h}$ is the position of the halo center for each halo, $\bm{x}_p$ is the position of the $p$-th member particle in the halo, $w_p(r_p)$ is the radial weight function and $r_p$ is the radius in the triaxial coordinate system defined by using the iterative scheme (see Appendix~\ref{app:definitions_of_inertia_tensor} for the details).
From the above consideration, we employ $w_p=1/r_p^2$; we upweight contributions from inner particles around the mass density maximum, assuming that those particles are more gravitationally bound and are proxies of stars if a galaxy forms in the halo \citep[see][for the similar discussion]{2013MNRAS.433.3506M}.

Taking $x^3$-direction as the line-of-sight direction, we define two components to characterize the ellipticity of each halo, from the inertia tensor, as
\begin{equation}
	\epsilon_+ \equiv \frac{I_{11} - I_{22}}{I_{11} + I_{22}},~\epsilon_\times \equiv \frac{2I_{12}}{I_{11} + I_{22}}.
	\label{eq:def_e}
\end{equation}
In an actual observation, we can see only the ``projected'' distribution of stars in each galaxy, and therefore the above definition would be appropriate for the definition of the halo ellipticity or closer to what we can estimate from the light distribution of each galaxy. 
In Appendix~\ref{app:definitions_of_inertia_tensor}, we study how the IA power spectra \mtrv{vary}
if 
\tkrv{different definitions of inertia tensors are used.}
A brief summary of the results is as follows. 
The ellipticities of individual halos are sensitive to how to define the 
\tkrv{inertia tensors}, and the large-scale IA amplitudes ($A_{\rm IA}$), measured from the simulations, also vary with the 
definitions. 
However 
\tkrv{the shape ($k$-dependence) of IA power spectrum and also the signal-to-noise ratio of it remain almost unchanged up to the scale sufficiently larger than the size of a halo ($k \lesssim 0.1~h{\rm Mpc}^{-1}$).
Hence as long as we marginalize $A_{\rm IA}$ as a free parameter like the linear halo/galaxy bias, we expect that the choice of shape measurement methods does not affect the results of a cosmological analysis with the IA power spectrum.} 

\begin{figure}
    \begin{center}
        \includegraphics[width=0.95\columnwidth]{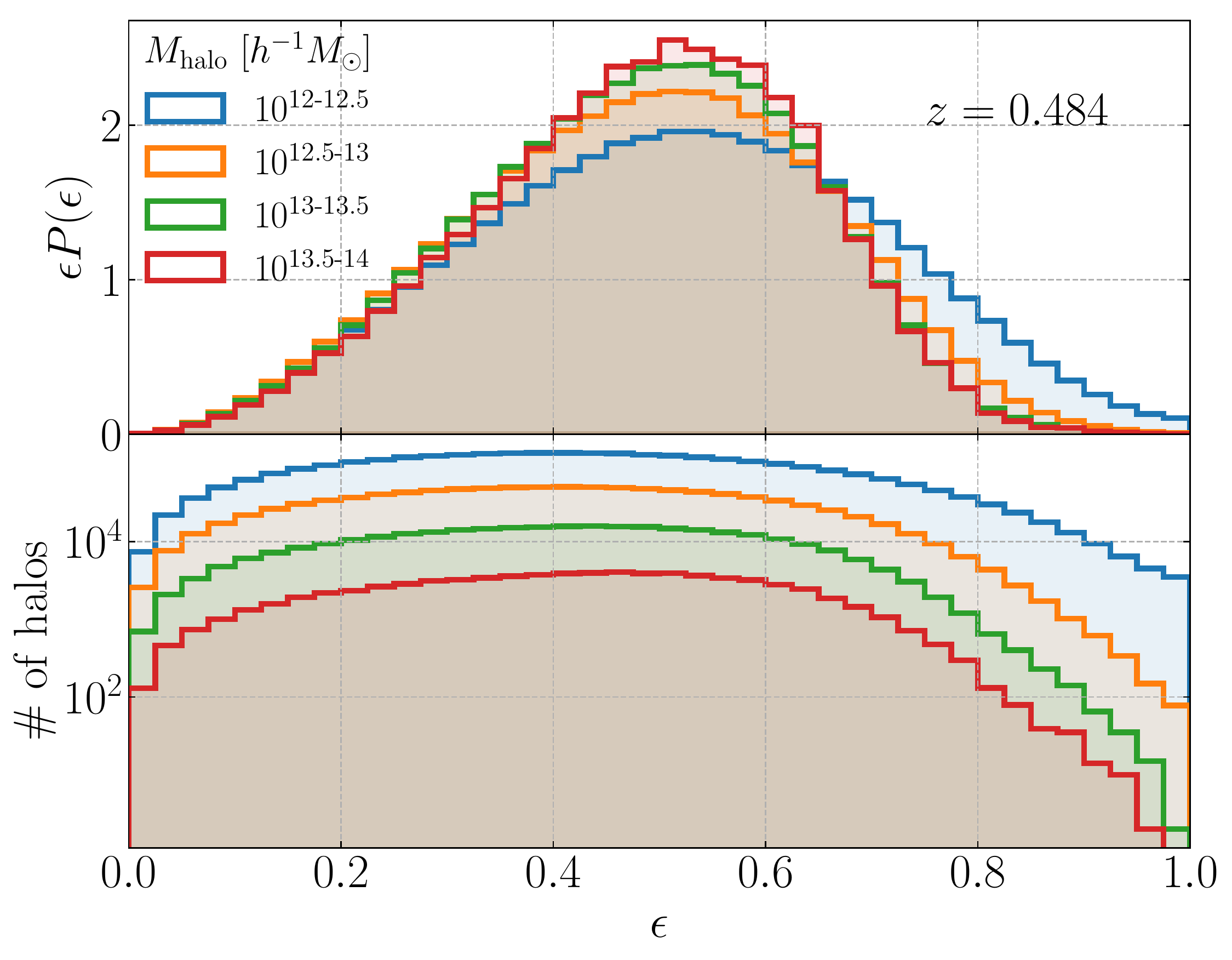}
    	\caption{Distributions of ellipticity amplitudes of halo shapes, defined as $\epsilon=\sqrt{\epsilon_+^2 + \epsilon_\times^2}$, for 
    	halo samples in the mass ranges denoted by the legend. 
    	Here we show the results at $z=0.484$. 
    	For the two-component ellipticities, the probability distribution is given as $P(\epsilon_+,\epsilon_\times)\mathrm{d}\epsilon_+ \mathrm{d}\epsilon_\times = P(\epsilon)\epsilon\mathrm{d}\epsilon\mathrm{d}\varphi$. 
    	Hence the top panel shows the probability distribution of ellipticity amplitude, $\epsilon P(\epsilon)$, which satisfies the normalization $\int_0^1\mathrm{d}\epsilon~ \epsilon P(\epsilon)=1$. 
    	The lower panel shows the non-normalized distribution, i.e., the number of halos, at each bin of $\epsilon$.
    	For illustrative purpose, we adopt the logarithmic scale of $y$-axis.
    	}
    	\label{fig:e_pdf}
	\end{center}
\end{figure}
Fig.~\ref{fig:e_pdf} shows the distribution of halo ellipticities measured from one simulation realization, for different halos samples, defined according to the halo mass ranges. 
Note that the distribution satisfies the normalization condition: $\int_0^1\mathrm{d}\epsilon~\epsilon P(\epsilon)=1$. 
The figure shows $\epsilon \sim 0.5$ as typical halo ellipticities, with a wide distribution. 
However, as we will show later, the IA effect arises from a correlated part between shapes of different halos, which corresponds to a few percent in the ellipticity amplitude, much smaller than the random intrinsic shapes. 
Thus the random intrinsic shapes give a dominant source of statistical errors in a measurement of the IA effect.
\tkrv{Note that the relatively larger ellipticity of low mass halos ($\sim 10^{12}~h^{-1}M_{\odot}$) is partly due to 
a finite number of member particles in individual halos, because
they contain only $\sim 100$ $N$-body particles, which is not enough to precisely measure the underlying shape of a halo and then 
adds statistical errors in 
the measured ellipticity per component; 
$\epsilon_{\rm rms} \equiv \sqrt{\langle \epsilon_+^2 \rangle } = \sqrt{\langle \epsilon_\times^2 \rangle }$ 
where $\langle \epsilon_+^2\rangle \equiv \frac{1}{N_h}\sum_h \epsilon^2_{+,h}$.
Nevertheless we find that the IA power spectra \mtrv{are not sensitive to the resolution issue due to a finite number of member particles in halos, 
as we will explicitly show in}
Appendix~\ref{app:particle_resolution}. }

\tkrv{
As in the weak lensing convention \citep{2002AJ....123..583B}, 
we convert the halo ellipticities into the 
shear, via the following relation, in order to compare with the IA theory given in terms of the gravitational tidal field:
\begin{equation}
    \gamma_{(+,\times)} = \frac{1}{2\mathcal{R}} \epsilon_{(+,\times)},
    \label{eq:gamma_N}
\end{equation}
where $\mathcal{R} \equiv 1-\epsilon^2_{\rm rms}$ is the responsivity \citep{2002AJ....123..583B}.
}
Typically $\mathcal{R}\simeq 0.9$ for our halo samples as indicted in Fig.~\ref{fig:e_pdf}.

\tkrv{
\subsection{Measurements of the IA power spectra}
\label{subsec:power_measurement}
}
The halo shape, given by Eq.~(\ref{eq:Iij}), is considered as a representative tracer of the underlying ellipticity/shear field or theoretically the tidal field in the IA model, which has an analogy  to the peculiar velocity field of galaxies \citep{Kaiser1987} or the weak lensing field \citep{2000MNRAS.313..524V}\footnote{We should keep in mind that the underlying tidal field is sampled at particular positions, i.e., halo positions, like the peculiar velocity field of galaxies \citep{2011JCAP...11..039S}.} for the definition.
In this work, we consider a number-density weighted ellipticity field:
\begin{equation}
	\tkrv{\hat{\gamma}}_{(+,\times)}(\bm{x}) = \frac{1}{\bar{n}_{\rm h}}\sum_{\rm h} \gamma^{\rm h}_{(+,\times)} 
	\delta_{\rm D}\!(\bm{x}-\bm{x}_{\rm h}),
	\label{eq:gamma_estimator}
\end{equation}
where $\bar{n}_{\rm h}$ is the mean number density of halos, and $\bm{x}_{\rm h}$ denotes the position of halos. 
It is useful to get access to this field on regular grids to make use of the Fast Fourier Transform.
To do so, we use the cloud-in-cell (CIC) assignment \citep{hockney81} to interpolate the ellipticities sampled at the positions of halos, to the entire simulation box (see Appendix~\ref{app:density_weighted_field} for details).
Throughout this paper we employ $1024^3$ grids to define the halo shear fields.
Finally we perform a Fourier-transformation of the fields to compute the $E$/$B$-mode fields from Eqs.~(\ref{eq:gammaE}) and (\ref{eq:gammaB}), $E(\bm{k})$ and $B(\bm{k})$.
After the decomposition, we measure the power spectrum from each realization; in the next section we consider the following power spectra: 
\begin{equation}
    \left\{P_\delta(k), P_{\rm h}(k), P^{(\ell)}_{\delta E}(k), P^{(\ell)}_{{\rm h}E}(k), P^{(\ell)}_{EE}(k)\right\}, 
\end{equation}
where ``$\delta$'' and ``${\rm h}$'' denote the density fields of matter and halos, respectively. 
We also give a discussion on the $B$-mode power spectrum in Appendix~\ref{app:shape_noise}.

\section{Results}
\label{sec:results}

\subsection{Power spectra of matter, halos and shapes}
\label{subsec:results_powerspectra}
\begin{figure}
    \begin{center}
    	\includegraphics[width=0.95\columnwidth]{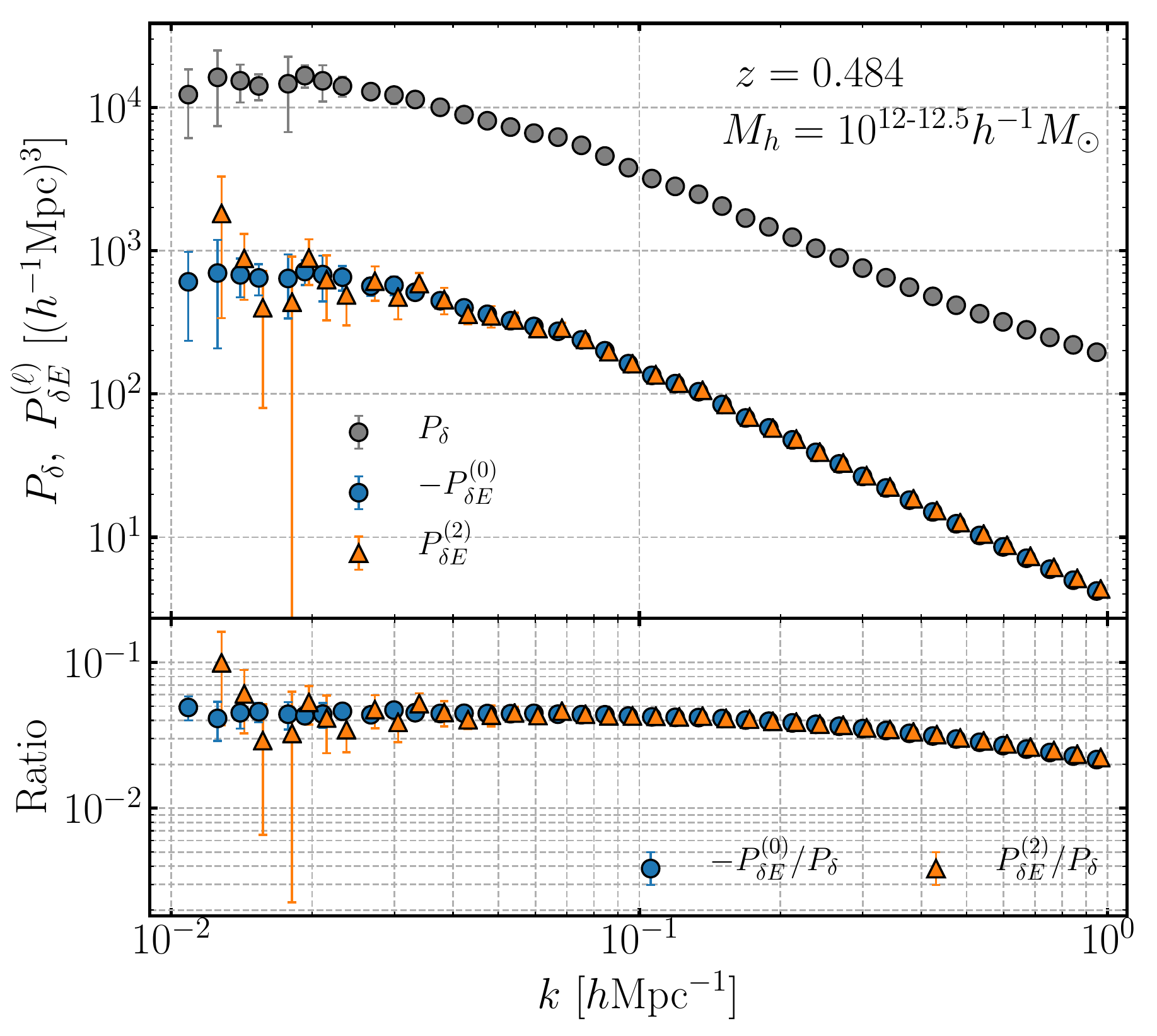}
    	\caption{The blue (orange) points in the top panel show the monopole (quadrupole) moment of the cross-power spectrum between the $E$ mode of halo shapes and the matter density field, $P_{\delta E}$, for the halo sample with $M_{\rm vir} = 10^{12-12.5}~h^{-1}M_{\odot}$ at $z=0.484$.
    	For comparison, the gray shows the ordinary matter power spectrum. 
    	The lower panel shows the ratios of the monopole or quadrupole moment to the matter power spectrum. 
    	}
    	\label{fig:Pds}
    \end{center}
\end{figure}
In Fig.~\ref{fig:Pds} we show the cross-power spectrum between the $E$-mode field of halo shapes and the matter density field, $P_{\delta E}(\bm{k})$, measured for halos with masses in the range $[10^{12},10^{12.5}]~h^{-1}M_\odot$ in simulation outputs at $z=0.48$. 
The symbols are the average among the 20 realizations, and the errorbars indicate the statistical error for a volume of $1~(h^{-1}{\rm Gpc})^3$, computed from the realization-to-realization scatters. 
First, the cross-power spectrum displays significant correlations over all the scales shown here, from the linear to nonlinear regimes, meaning that halo shapes have a physical correlation with the surrounding matter density field on all scales beyond a size of halos $(\sim h^{-1}{\rm Mpc}$ at most) as predicted by the tidal alignment model. 
Reflecting the spin-2 field nature of halo shapes, the cross-power spectrum has both the monopole and quadrupole moments. 
However, the relation between the two moments is purely geometrical, and the simulation result confirms that $P^{(0)}_{\delta E}(k)=-P^{(2)}_{\delta E}$ holds even at nonlinear scales (high $k$ bins beyond $k\simeq 0.1$~\hMpci).
The minus sign of the monopole moment indicates that halo shapes are stretched in the direction of the minor axis of the tidal field, which means that the principal major axis of a halo's inertia tensor tends to be aligned with the filament structure or on the sheet structure for instance.
\begin{figure}
    \begin{center}
    	\includegraphics[width=0.95\columnwidth]{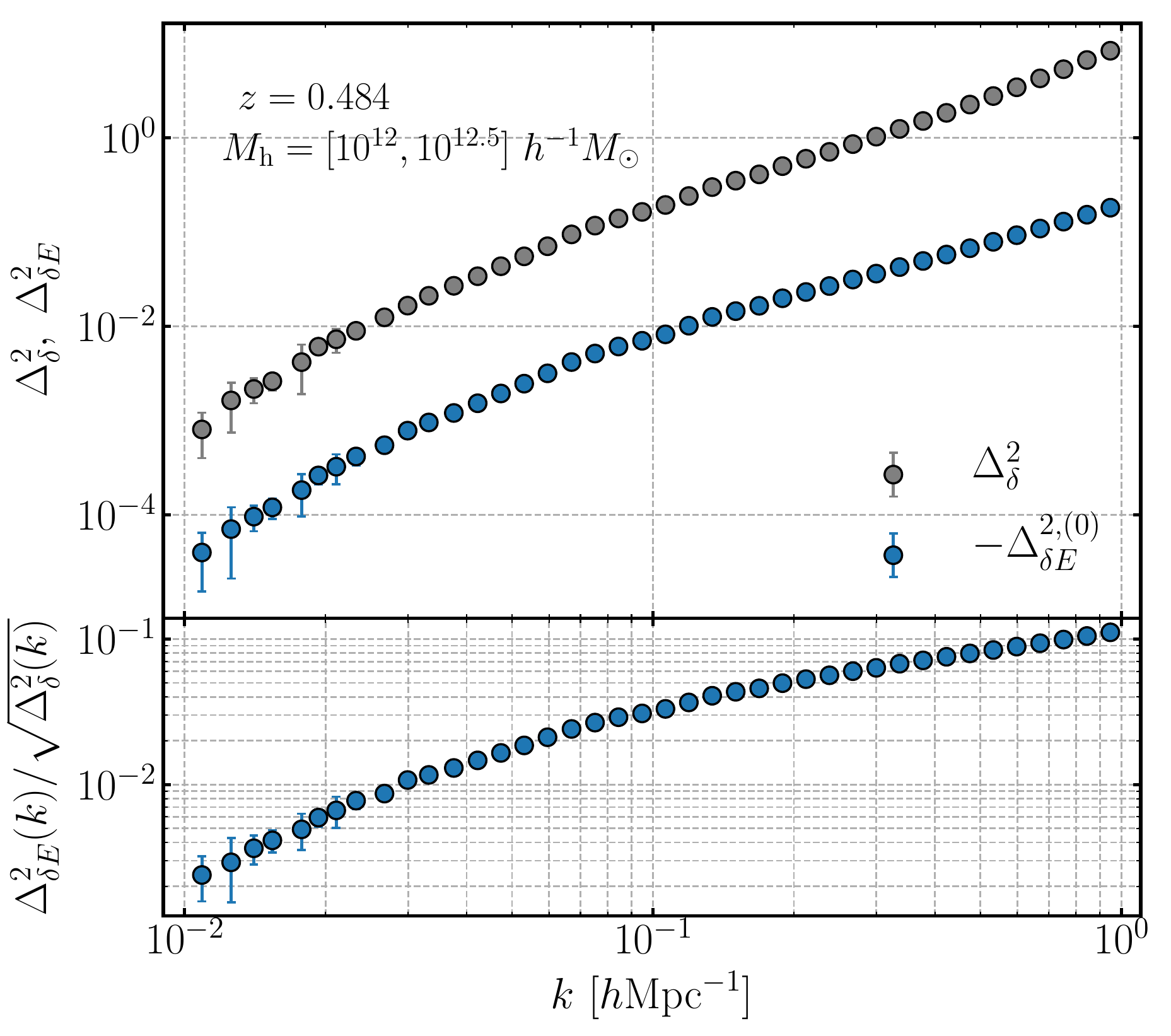}
    	\caption{Similarly to the previous figure, but shown is the dimensionless power spectra, defined as $\Delta^2_\delta(k)\equiv k^3P_\delta(k)/(2\pi^2)$ and $\Delta^2_{\delta E}(k)\equiv k^3P_{\delta E}(k)/(2\pi^2)$, respectively. 
    	The dimensionless power spectrum at a particular wavenumber $k$ gives an amplitude of the real-space variance at the corresponding wavelength; 
    	e.g., $\Delta_{\delta E}(k)^2\simeq \left.\langle\delta\gamma_E\rangle\right|_{\lambda\sim 1/k}$ in the linear or weakly nonliner regime (see text for details). 
    	The lower panel shows an amplitude of the real-space IA shear at the wavelength, $\Delta^2_{\delta E}(k)/[\Delta^2_{\delta}(k)]^{1/2}\sim \left.\gamma_E\right|_{\lambda\sim 1/k}$ in the linear regime.}
    	\label{fig:EmodeAmp}
    \end{center}
\end{figure}

The lower panel of Fig.~\ref{fig:Pds} shows the ratio of $P_{\delta E}(k)$ to $P_{\delta}(k)$. 
The ratio approaches a constant value at the limit $k\to 0$. 
This asymptotic behavior is analogous to a linear bias coefficient, e.g., as seen from the ratio of the halo-matter cross power spectrum to the matter power spectrum, $P_{\delta {\rm h}}/P_{\delta}= b_{\rm h}$ at $k\to 0$ with a constant coefficient $b_{\rm h}$. 
The scale-independent (constant) ratio is a confirmation of the linear alignment model. 
This large-scale correlation is as expected in the standard $\Lambda$CDM model with an adiabatic Gaussian initial condition that is employed in our simulations, as follows. 
The formation and evolution of individual halos are governed by local physics or physical quantities within a few Mpc scales around each halo. 
Hence, as long as the physical correlation of halo shapes with the large-scale matter distribution arises on scales beyond the halo scales, it should originate from the gravitational interaction and the primordial perturbations. 
Since there is only a single degree of freedom in the perturbations at large scales in the adiabatic initial conditions, the power spectra of the IA (halo shape) fields at linear scales should be related to the matter power spectrum via a constant factor \citep[also see][for the similar discussion on halo bias]{2018PhR...733....1D}. 
The small-$k$ constant ratio of Fig.~\ref{fig:Pds} indicates that halo shapes retain the information on the primordial density perturbation on large scales, very similarly to what the density perturbation of halos does.

In Fig.~\ref{fig:EmodeAmp}, we show the dimensionless power spectra, defined by $\Delta^2_{XY} \equiv k^3P_{XY}/2\pi^2$, to study the typical amplitude of the halo shape $E$-mode field. 
Recalling that the dimensionless power spectrum at a particular $k$ corresponds to the real-space variance per unit logarithmic wavenumber interval at the corresponding length scale, e.g., $\Delta^2_\delta(k)\sim \left.\langle \delta^2_R\rangle\right|_{R\sim 1/k}$, one can find $\delta\sim O(1)$ at $k\sim$a few $O(0.1)$~\hMpci from the gray points showing $\Delta^2_\delta$. 
Then comparing the amplitudes of $\Delta^2_\delta$ and $\Delta^2_{\delta E}$ tells $E\sim$a few $O(0.01)$, i.e., a few percent for the IA shear amplitude at $k\simeq 0.1$~\hMpci. 
This means that, if the halo $E$-mode field is smoothed within a volume of scales $R_{\rm sm}\sim 1/k\sim \mbox{a few 10~Mpc}$, the $E$-mode amplitude is of the order of 0.01.
This $E$-mode amplitude can be compared to the intrinsic random shape, $\gamma_N \sim 0.2$ (Fig.~\ref{fig:e_pdf} and Eq.~\ref{eq:gamma_N} taking into account the relation $\gamma=\epsilon/(2{\cal R})$ with responsivity ${\cal R}\sim 0.9$). 
Thus the large-scale IA shear is measurable only in a statistical sense, e.g., via the correlation function or power spectrum for the two-point statistics. 
At the nonlinear scales $k\gtrsim 0.1$~\hMpci, the shear IA amplitude appears greater as shown by the lower panel, but the boosted amplitudes are likely due to the higher-order contribution of density perturbation as explained around Eq.~(\ref{eq:gamma_weighted}). 

\begin{figure}
    \begin{center}
    	\includegraphics[width=0.95\columnwidth]{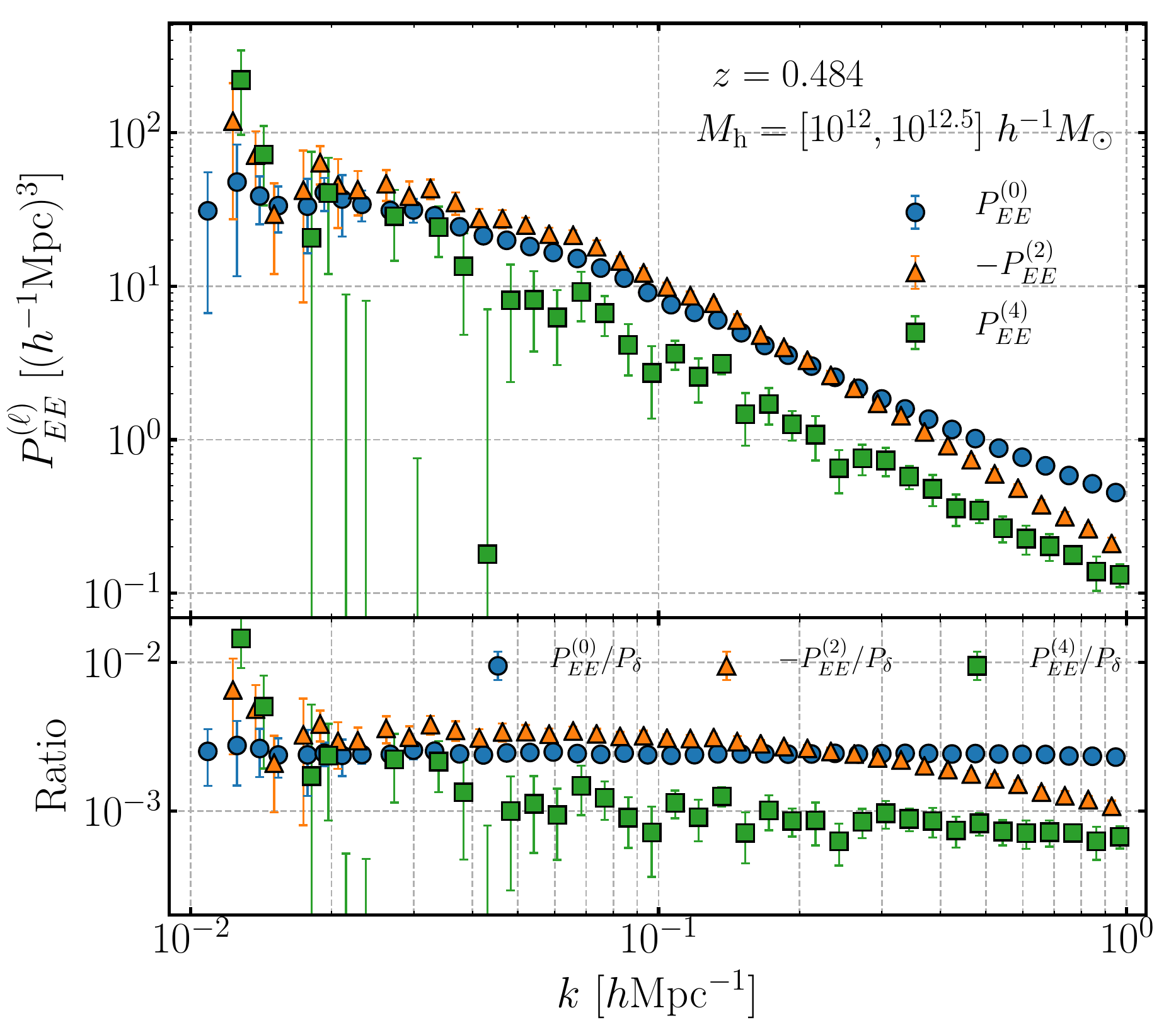}
    	\caption{The blue, orange, and green data points show the monopole, quadrupole and hexadecapole moments of the auto power spectrum of the $E$ mode, respectively, for the same halo sample as in Fig.~\ref{fig:Pds}. 
    	We subtracted the shot noise term from the measured power spectrum (see Appendix~\ref{app:shape_noise} for details).}
    	\label{fig:Pss}
    \end{center}
\end{figure}
In Fig.~\ref{fig:Pss} we show the auto-power spectra of the $E$-mode shape field. 
Here we first subtracted the shot noise term from the measured power spectrum, and then computed the multipole moments of power spectrum. 
In Appendix~\ref{app:shape_noise} we in detail describe how to estimate the shot noise term due to the discrete nature of the intrinsic shapes of halos in each simulation. 
Note that the shape noise contributes only to the monopole moment.
The monopole and quadrupole moments display different $k$-dependences at $k\gtrsim 0.1~h~{\rm Mpc}^{-1}$. 
This means that a simple geometrical relation between the monopole and quadrupole moments, given by Eq.~(\ref{eq:relations_moments}), does not hold for the auto spectrum especially at $k\gtrsim 0.1~$\hMpci, unlike that of the cross-power spectra (the relation for the \tkrv{hexadecapole} moment is not clear due to the larger errors).  
This implies that the higher-order contributions to the auto-power spectra cause non-trivial angular modulations, which are also found from a perturbation theory calculation in \citet{Blazeketal2015}.
In Appendix~\ref{app:shape_noise}, we show that the $B$-mode auto-power spectrum displays a deviation from the simple shot noise, with a weak-scale dependence (see Fig.~\ref{fig:shapenoise}). 
We believe that this is ascribed to  the ``renormalized'' shot noise arising from the small-scale nonlinear terms as discussed in \citet{Blazeketal2017}, which has an analogy to the renormalization of bias parameters \citep{2006PhRvD..74j3512M,2009JCAP...08..020M}. 
This term should equally contribute to the $E$-mode auto-power spectrum. 

The IA effect is one of the most important systematic effects in cosmic shear cosmology \citep{2017MNRAS.465.1454H,2018PhRvD..98d3528T,2019PASJ...71...43H,2020PASJ...72...16H}. 
In this context, there are two contributions, called ``II'' and ``GI'', to the cosmic shear power spectrum, which correspond to $P_{EE}$ and $P_{\delta E}$, respectively. 
In cosmic shear analyses, the following relation is often {\it assumed} based on the linear alignment model (Eq.~\ref{eq:def_of_AI}): 
\begin{align}
    P_{\delta E}(k)&= - A_{\rm IA}c(z)(1-\mu^2)P_\delta(k),\nonumber\\
    P_{EE}(k)&= A_{\rm IA}^2c(z)^2 (1-\mu^2)^2 P_\delta(k), 
\label{eq:NLA}
\end{align}
where $c(z)$ is a 
\tkrv{scale-independent factor}
at a particular redshift, defined from Eq.~(\ref{eq:def_of_AI}) as $c(z)\equiv C_1\rho_{\rm cr0}\Omega_{\rm m}/D(z)$.
The cosmic shear is a projected field of the underlying matter density field along the line-of-sight direction, so the power spectrum corresponds to the one evaluated at $\mu=0$ in the above equation because $k_\perp = k(1-\mu^2)^{1/2}$ (see Section~\ref{subsec:2d} for a similar discussion).
If we use the nonlinear power spectrum for $P_\delta(k)$, it corresponds to the nonlinear alignment (NLA) model \citep{2007NJPh....9..444B}.

\tkrv{
\begin{figure*}
    \begin{center}
        \includegraphics[width=1.9\columnwidth]{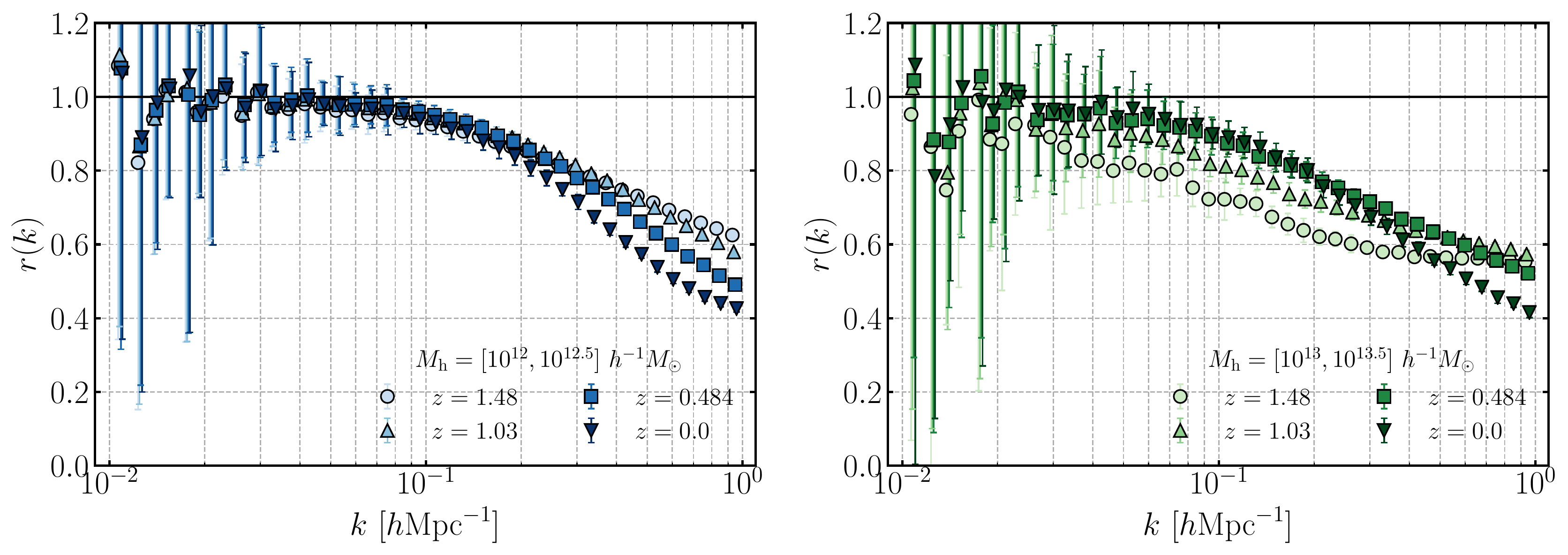}
        \caption{The cross-correlation coefficients $r(k)$ of the IA cross-power spectrum, $P^{(0)}_{\delta E}$ for various halo masses and redshifts (see Eq.~\ref{eq:rcc} for the definition of $r$). If the linear alignment model (Eq.~\ref{eq:NLA}) holds, 
        this coefficeint $r=1$. 
        }
        \label{fig:corrcoeff}
    \end{center}
\end{figure*}
Here we address the validity of the linear and nonlinear alignment models by comparing the expressions (Eq.~\ref{eq:NLA}) with the IA power spectra measured in simulations. 
Fig.~\ref{fig:corrcoeff} shows the correlation coefficient of the matter density field and $E$-mode field defined as:
\begin{equation}
    r^2(k) \equiv \mathcal{N}\frac{\left[P^{(0)}_{\delta E}(k)\right]^2}{P_\delta(k) P^{(0)}_{E E}(k)},
    \label{eq:rcc}
\end{equation}
where the prefactor $\mathcal{N} = 6/5$ normalizes $r(k)$ as unity if the linear or nonlinear alignment model holds\footnote{\tkrv{Since the IA power spectrum is intrinsically anisotropic, the correlation coefficient should be a function of $k$ and $\mu$; $r^2(k,\mu) \equiv P^2_{\delta E}(k,\mu) / (P_\delta(k) P_{EE}(k,\mu))$. The definition of Eq.~(\ref{eq:rcc}) then only focuses on the monopole component of the cross-correlation.
Indeed, we obtained similar results with Fig.~\ref{fig:corrcoeff} from the measured $r^2(k,\mu)$ with fixed $\mu$.}}. 
}
\tkrv{
In the $k\to0$ limit, $r$ goes to unity, i.e. the linear alignment model is valid at each redshift for the low mass samples.
Note that it apparently does not hold for the high mass samples at high redshifts due to the non-Poisson shape noise; 
since we here only subtract the pure Poisson shape noise from the measured $P^{(0)}_{EE}$, it still has the non-Poissonian contribution in small $k$-bins. 
That positive residual offset causes $r<1$ at small $k$. 
We checked $P^{(0)}_{EE}$ is actually well fitted by the linear alignment model after subtracting the non-Poisson shape noise which is estimated by $P^{(0)}_{BB}$ (see Apprendix~\ref{app:shape_noise}). 
}
\tkrv{\mtrv{On the other hand, at nonlinear scales $k\gtrsim 0.1~h{\rm Mpc}^{-1}$ the cofficient $r$ goes below unity,}
and the IA power spectra display different shapes from the alignment models in the nonlinear regime.
If we recall that most of cosmological information in the cosmic shear power spectrum are from the scales in $k\simeq [0.1,1]~h~{\rm Mpc}^{-1}$ \citep{2005APh....23..369H,2017MNRAS.465.1454H,2018PhRvD..98d3528T,2019PASJ...71...43H,2020PASJ...72...16H}, the violation of the relation (Eq.~\ref{eq:NLA}) might cause a bias in the cosmological parameters, derived by marginalizing over the IA parameters. 
Therefore, the potential impact of this breakdown of the commonly-used model should be carefully studied.
}

\subsection{Mass and redshift dependences of $A_{\rm IA}$}
\label{subsec:results_halomassetc}

\begin{figure*}
    \begin{center}
        \includegraphics[width=0.95\columnwidth]{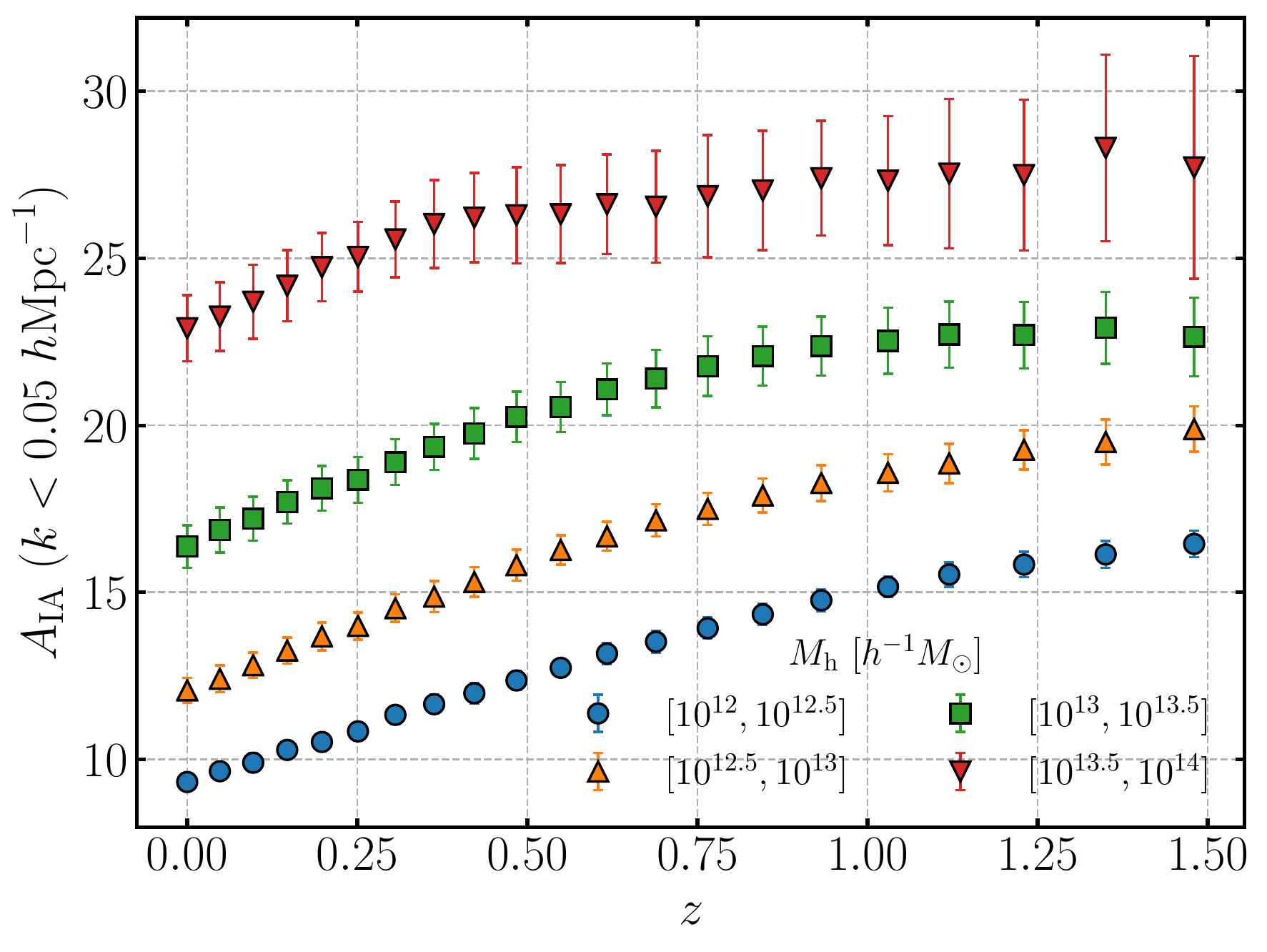}
        \hspace{5pt}
        \includegraphics[width=0.95\columnwidth]{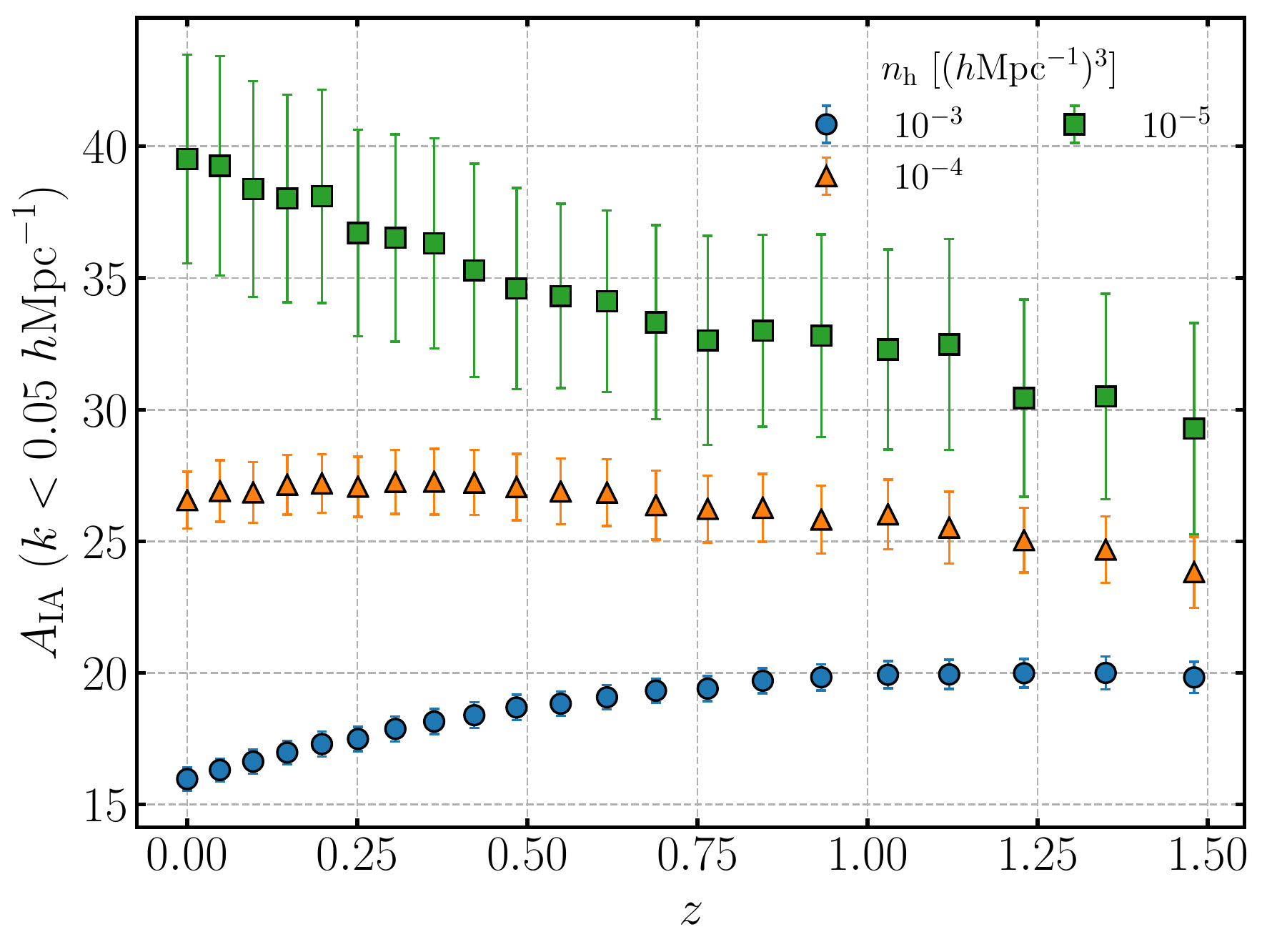}
    \end{center}
    \caption{The halo mass dependence and the redshift evolution of the large-scale IA coefficient, $A_{\rm IA}$, which is estimated according to Eq.~(\ref{eq:AI_est}) and theoretically corresponds to the coefficient of linear alignment model in Eq.~(\ref{eq:def_of_AI}). 
    {\em Left panel}: The results for the samples of halos in a given mass range, as denoted by the legend. 
    {\em Right}: Similar to the left panel, but the results for samples of halos with a fixed number density, where we define each sample by selecting halos from the ranked list of masses starting from the most massive halo until the number density of selected halos matches the target number density.}
    \label{fig:amp}
\end{figure*}
In Fig.~\ref{fig:amp} we study how the linear IA coefficient, $A_{\rm IA}$ (see Eq.~\ref{eq:def_of_AI}) varies with redshift and halo mass.
We estimate $A_{\rm IA}$ by minimizing the following $\chi^2$ with varying a parameter $\hat{A}_{\rm IA}$,
\begin{equation}
    \chi^2\equiv \sum_{k_i; k_i<0.05~h~{\rm Mpc}^{-1}}
    \frac{[R(k_i)-(2/3)c(z)\hat{A}_{\rm IA}]^2}{\sigma_{R_i}^2},
    \label{eq:AI_est}
\end{equation}
where $c(z)$ is the same 
\tkrv{factor} defined below Eq.~(\ref{eq:NLA}), $R(k_i)$ is the ratio of the monopole of matter-IA cross-power spectrum to the matter power spectrum in the $i$-th $k$ bin, defined as $R(k_i)\equiv P^{(0)}_{\delta E}(k_i)/P_{\delta}(k_i)$, a factor of $2/3$ in front of $c(z)$ is from the $\mu$-integral of $(1-\mu^2)$ in the monopole calculation of $P^{(0)}_{\delta E}$, and $\sigma_{R_i}^2$ is the variance of the ratio in the $k$-bin, estimated from the 20 simulation realizations.
Here we consider two sets of halo samples with different selection rules; one set is a halo sample selected in a given mass range (mass-bin sample), while the other is specified by a fixed number density of halos. 
For the latter, we select halos from the ranked list of masses starting from the most massive one at each redshift output until the number density of selected halos matches the target value.
Note that the mass-bin samples have different number densities at different redshifts.
The figure shows the best-fit coefficients $A_{\rm IA}$ for each halo sample at a given output redshift.
We should again remind that $A_{\rm IA}$ is, by construction, defined with respect to the primordial gravitational potential (or curvature) perturbations at large scales (small $k$'s), which are constant in time.  
We begin with  the results for the mass-bin samples, which show several interesting trends. 
First, the figure shows that $A_{\rm IA}$ is greater for more massive halos at a fixed redshift. 
Second, $A_{\rm IA}$ is greater at higher redshifts for a fixed mass-bin halo sample. 
These results reflect that more massive halos and halos at higher redshift have a greater response to the linear tidal field.
Third, the $A_{\rm IA}$ values for the two samples at the high-mass end (red and green points in the left panel) show a plateau, approaching to an asymptotic constant value in high redshift bins, as predicted by the linear IA model arising from the primordial tidal field that is constant in time (therefore leading to a time-independent $A_{\rm IA}$)\footnote{If the linear alignment model (Eq.~\ref{eq:la}) holds, the linear IA coefficient ($A_{\rm IA}$) for halos of the same mass would be the same and constant in time, whenever the IA correlation is measured (even if the abundance of the halos significantly changes across different redshifts). 
This is because the halos of same mass form from the primordial density peaks of the same Lagrangian volume and the large-scale  relation/correlation between the halo shapes and the primordial tidal field has no time dependence in the Lagrangian picture.}.
\mtrv{We checked that the halo mass dependence of $A_{\rm IA}$, especially before the plateau, is qualitatively consistent with the result in \citet{2018MNRAS.474.1165P}, which found that the linear IA coefficient scales with halo mass as $A_{\rm IA}\propto M_{\rm h}^{\beta}$ with $\beta\simeq 0.3$ for $M_{\rm h} \simeq [10^{13},10^{14}]~h^{-1}M_\odot$ from the 
Millenium simulation.}

Now we consider the samples for a fixed number density. 
A spectroscopic survey of galaxies is sometimes designed to keep a constant number density over a range of redshifts for the cosmological analysis purpose \citep[e.g.,][]{2013AJ....145...10D,2014PASJ...66R...1T}. 
The ongoing and upcoming spectroscopic surveys are in the range of $\bar{n}=[10^{-4},10^{-3}]~(h~{\rm Mpc}^{-1})^3$.
The redshift evolution of $A_{\rm IA}$ depends on the number density of a sample; $A_{\rm IA}$ decreases with the increase of redshift for a low density sample such as $\bar{n}= 10^{-5}~(h~{\rm Mpc}^{-1})^3$, $A_{\rm IA}$ appears to be almost constant with respect to redshifts for $10^{-4}~(h~{\rm Mpc}^{-1})^{-3}$ and $A_{\rm IA}$ increases with redshift for $10^{-3}~(h~{\rm Mpc})^{-3}$. 
Thus the $A_{\rm IA}$ amplitude depends on the selection of halos or the nature of the halo sample. 
Finally, we comment on a connection of the results in Fig.~\ref{fig:amp} to the IA effects of galaxies. 
We can consider the $A_{\rm IA}$ amplitude shown in Fig.~\ref{fig:amp} is the maximum case, since we consider the halo shapes. 
Since the physics and evolution of galaxies are more complicated, and galaxy shapes would have a misalignment with the halo shapes to some degrees \citep{Okumuraetal2009}, the $A_{\rm IA}$ coefficients for galaxies would be smaller even if the galaxies of interest are central galaxies and reside in halos in the mass range we have considered so far. 
We also note that the $A_{\rm IA}$ amplitude varies with the definition of halo shapes even for the same sample of halos, as shown in Appendix~\ref{app:definitions_of_inertia_tensor}.

Fig.~\ref{fig:amp} indicates $A_{\rm IA}\sim 20$ for halos with $10^{13}~h^{-1}M_\odot$ at $z\sim 0.5$,
which roughly corresponds to the host halos of the SDSS luminous red galaxies, and this is larger than $A_{\rm IA}\sim 8$ implied from the actual SDSS data \citep{Okumuraetal2009,Singhetal2015}. 
As discussed in Appendix~\ref{app:definitions_of_inertia_tensor}, if we employ a crude definition of the halo shapes in simulations, it leads to about halved value of $A_{\rm IA}$  even for the same sample of halos\footnote{Nevertheless, note that, even for this case, the signal-to-noise ratios of IA power spectrum is not largely changed as in the main results we show below.}.
In addition, actual galaxies might have a misalignment with the orientations of the host halos, and this also leads to a smaller $A_{\rm IA}$ value inferred from galaxy shapes, compared to the halo shapes \citep{Okumuraetal2009}. 
A random misalignment of about 30~degrees between the major axes of halo and galaxy orientations leads to about factor of 2 smaller value of $A_{\rm IA}$. 
Thus an actual value of $A_{\rm IA}$ is sensitive to the definition of halo shapes and the properties of galaxies relative to host halos, so the results of Fig.~\ref{fig:amp} can be considered as an example of $A_{\rm IA}$ values that dark matter halos could have. 
Or the parameter $A_{\rm IA}$ should be considered as a ``nuisance''  parameter, because the genuine value of $A_{\rm IA}$ is difficult to predict from the first principles.

\tkrv{
In Appendix~\ref{app:b_K}, we show the mass and redshift dependence of another definition of the linear coefficient, $g_{ij} \equiv b_K K_{ij} $, which is commonly used in the context of the perturbation theory of the IA physics for convenience.
}

\subsection{Baryon Acoustic Oscillation features}
\label{subsec:bao}
\begin{figure*}
	\includegraphics[width=1.9\columnwidth]{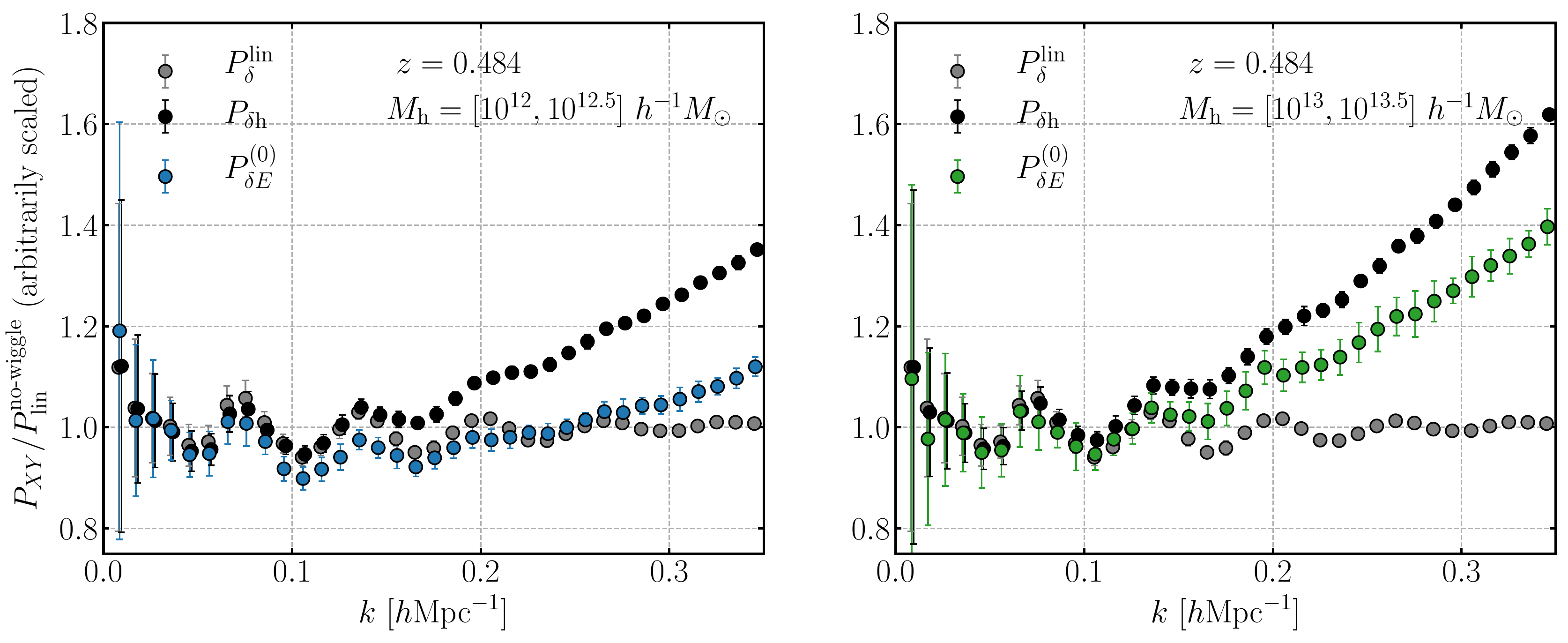}
	\caption{The ratio of the cross-power spectra of matter and halo $E$-field to the linear matter power spectrum without BAO wiggles (no wiggle), for halos samples in different mass bins at $z=0.484$. 
	Here we use the fitting formula in \citet{1998ApJ...496..605E} to compute the no-wiggle matter spectrum for the {\it Planck} cosmology, the same model used in the simulations. 
	For easier comparison we arbitrarily normalize the ratio in such that all the curves have the similar amplitudes up to $k=0.05~h~{\rm Mpc}^{-1}$ (we employed the normalization on individual realization basis).
	We here consider the two samples of halo masses, where the sample of $M_{\rm h}=[10^{13},10^{13.5}]h^{-1}M_\odot$ roughly corresponds to a typical mass of halos hosting the BOSS CMASS galaxies. 
	We also show the ratios for the matter-halo cross spectrum, $P_{\delta{\rm h}}$, and for the linear matter power spectrum with BAO wiggles, which are similarly normalized (arbitrarily scaled in the $y$-directoin).
	}
	\label{fig:bao}
\end{figure*}
In Fig.~\ref{fig:bao} we show the ratio of the cross-power spectrum of matter and halo shapes to the linear ``no-wiggle'' matter power spectrum for several halo samples in different mass bins, where we use \citet{1998ApJ...496..605E} to compute the linear matter spectrum with no BAO features for the {\it Planck} cosmology.
We arbitrarily normalize all the cross-power spectra so that the ratio, $P_{\delta E}(k)/P_{\delta}(k)$, is close to unity at $k$ bins up to $k=0.05~h~{\rm Mpc}^{-1}$ in each realization.

The IA power spectrum displays clear BAO features as in the power spectrum of the halo density field. 
Thus the IA power spectrum can be used to measure the BAO scales \citep{2019PhRvD.100j3507O}.
Perhaps more interestingly, while the power spectrum of the halo density field has a boost in the amplitude at $k\gtrsim 0.1~$\hMpci in the nonlinear regime, the IA power spectrum displays a weaker boost in the amplitude at such nonlinear scales; for less massive halos with $M_{\rm h}=[10^{12},10^{12.5}]~h^{-1}M_\odot$ the amplitude stays almost unchanged as that of the linear power spectrum. This could be interpreted as follows.
Consider an overdensity region in the initial linear density field, at a sufficiently high redshift, i.e., in the linear regime. 
The Lagrangian volume of such a region shrinks due to the gravitational instability, and the density contrast accordingly grows due to the mass conservation.
A larger number of halos form in such an overdensity region.
Thus the mass density or number density of halos have a boost in the amplitude, reaching $\delta\gg 1$, in the overdensity region. 
On the other hand, there is no conservation law for the halo shapes or tidal fields.
Even in the highly nonlinear regime, ellipticities of halo shapes still stay in the range of $|\boldsymbol{\gamma}|\le 1$ or never goes beyond unity, unlike the density contrast. 
Hence the IA power spectrum should have a weaker response to the nonlinear clustering, at least in the power spectrum amplitudes.
Nevertheless, the observed IA field is a galaxy density-weighted field (see below), and the observed halo shapes are expressed as $(1+\delta)\boldsymbol{\gamma}$. 
The prefactor $(1+\delta)$ can lead to a boost in the IA power spectrum, which partly explains a boost in the IA power spectrum for the halo sample with $M_{\rm h}=[10^{13},10^{13.5}]~h^{-1}M_\odot$. 
These are interesting results. 

\subsection{Signal-to-noise ratio}
\label{subsec:snr}
\begin{figure*}
	\includegraphics[width=1.9\columnwidth]{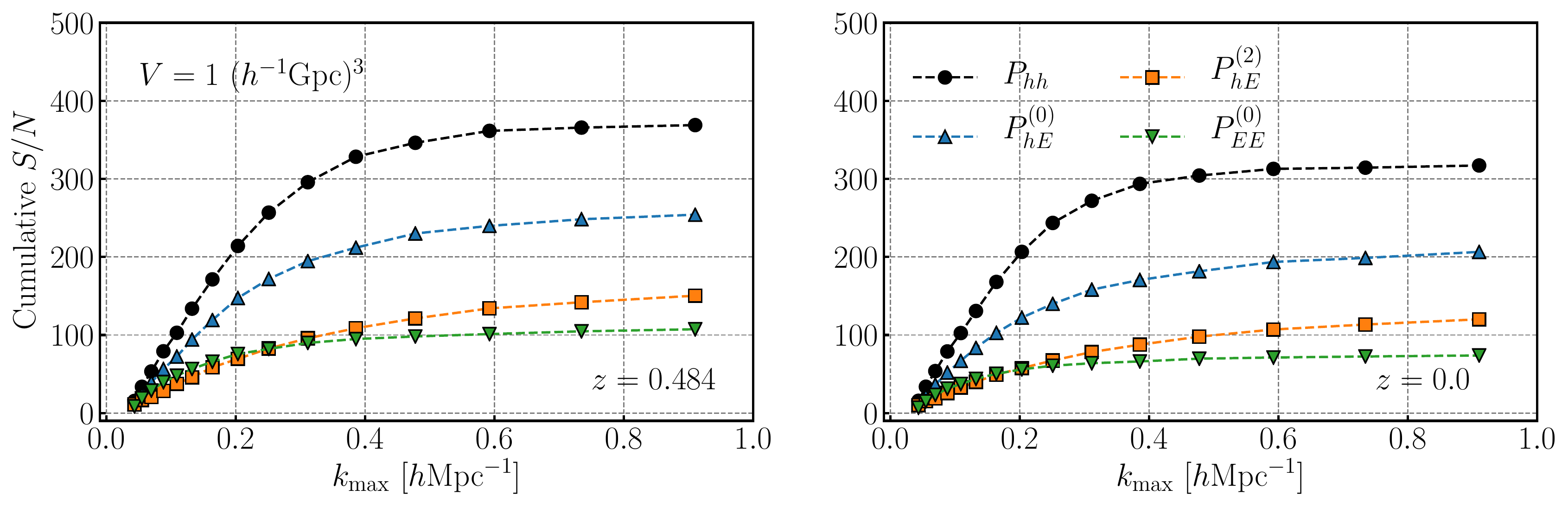}
	\caption{The cumulative signal-to-noise ($S/N$) for the halo power spectrum and
	the IA power spectra as a function of $k_{\rm max}$, where the $S/N$ is defined by integrating the differential $S/N$ over $0.04 \le k \le k_{\rm max}$ (see text for details). 
	Here we consider the halo sample with  $M_{\rm vir} = 10^{12\mathchar`-12.5}~h^{-1}M_\odot$ at two redshift outputs, $z=0.484$ (left panel) and $z=0.0$ (right), respectively. 
	The black, blue, orange and green data correspond to $P_{hh},P^{(0)}_{hE},P^{(2)}_{hE},P^{(0)}_{EE}$, respectively. 
	The results correspond to the $S/N$ values for a survey volume of $V=1~(h^{-1}~{\rm Gpc})^3$.
	}
	\label{fig:snr_all}
\end{figure*}
\begin{figure*}
	\includegraphics[width=1.9\columnwidth]{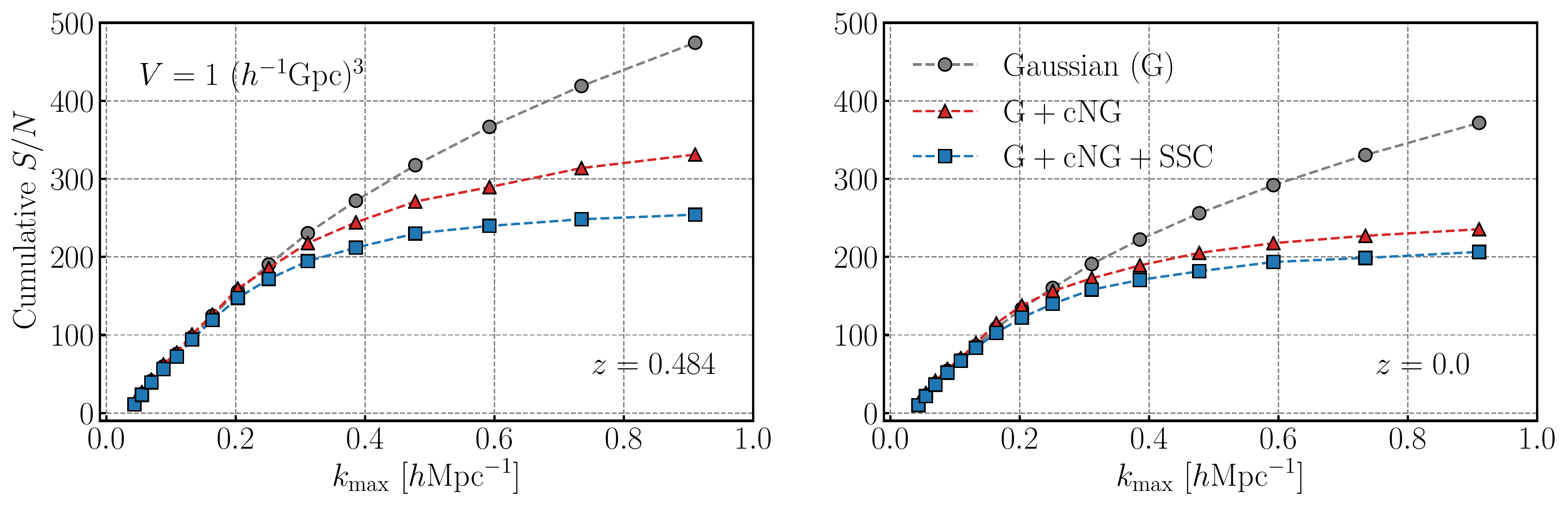}
	\caption{Similar to the previous figure, but shown is the relative importance of the Gaussian, the connected non-Gaussian and the SSC covariance contributions in the $S/N$ calculation for the monopole moment $P^{(0)}_{{\rm h}E}$.
	The blue data are the same one in Fig.~\ref{fig:snr_all} for the full covariance including the SSC terms, and the gray or red data points are the results when including the Gaussian covariance alone or ignoring the SSC contribution, respectively.
	}
	\label{fig:snr_ssc}
\end{figure*}

How much information does the IA power spectrum carry, compared to the standard halo power spectrum? 
To address this question, we study the cumulative signal-to-noise ratio ($S/N$) over a range of $k_{\rm min}\le k\le k_{\rm max}$, defined by
\begin{equation}
    \left(\frac{S}{N}\right)^2 \equiv  \sum_{k_i = k_{\rm min}}^{k_{\rm max}} \bar{P}^{(\ell)}(k_i) \left[{\bf C}_{(\ell \ell')}\right]^{-1}_{ij} \bar{P}^{(\ell')}(k_j),
    \label{eq:snr_ssc}
\end{equation}
where ${\bf C}_{(\ell\ell')}$ is the covariance matrix between the $\ell$- and $\ell'$-th multipole moments of power spectra and $[{\bf C}_{(\ell\ell')}]^{-1}$ is the inverse of the covariance matrix. 
Given an estimator of the power spectrum, the covariance matrix is defined as
\begin{align}
    {\bf C}_{(\ell\ell')ij}&\equiv \left\langle \hat{P}^{(\ell)}\!(k_i)\hat{P}^{(\ell')}\!(k_j)\right\rangle -\left\langle \hat{P}^{(\ell)}\!(k_i)\right\rangle \left\langle \hat{P}^{(\ell')}\!(k_j)\right\rangle \nonumber\\
    &= {\bf C}^{{\rm G}}_{(\ell\ell')ij}+{\bf C}^{{\rm cNG}}_{(\ell\ell')ij}+{\bf C}^{{\rm SSC}}_{(\ell\ell')ij},
    \label{eq:cov_ssc}
\end{align}
and $\bar{P}^{(\ell)}(k_i)\equiv \langle \hat{P}^{(\ell)}(k_i)\rangle$. 
Throughout this paper, we adopt $k_{\rm min}=0.04~h~{\rm Mpc}^{-1}$ for the minimum wavenumber and $\Delta {\rm ln}k = 0.215$ for the width of the $k$-bin in the $S/N$ calculation.
The covariance can be generally broken down into three contributions \citep{2013PhRvD..87l3504T}; the Gaussian (G) covariance, the connected non-Gaussian (cNG) covariance, and the super-sample covariance (SSC), respectively.
These covariance contributions to the IA power spectrum have not been studied. 
For the Gaussian field, the covariance has only the Gaussian contribution. 
The non-Gaussian covariances (cNG plus SSC) arise from the nonlinear mode coupling, more specifically the four-point correlation function (trispectrum) of the fields. 

To accurately estimate the covariance matrices of the halo and IA power spectra, we use a set of the simulation realizations following the method in \citet{2014PhRvD..89h3519L}. 
We use a suite of 20 simulations in \citet{2019ApJ...884...29N} each of which has a 1~$h^{-1}~{\rm Gpc}$ box size. 
We subdivide each box into 64 subvolumes of size $250~h^{-1}~{\rm Mpc}$ each.
Thus we have $N_{\rm sub}=20\times 64=1280$ subboxes in total. 
We measure the power spectrum, $\hat{P}_{XY}$, from each of the subboxes, and then take the the standard estimator to obtain the covariance of the sub-volume power spectra: 
\begin{align}
	{\bf C}_{(\ell\ell')ij} &\equiv \frac{1}{N_r-1} \sum_{\alpha=1}^{N_r}\left(\hat{P}^{(\ell)}_{XY,\alpha}(k_i) - \bar{P}_{XY}^{(\ell)}(k_i)\right) \nonumber\\
	&\hspace{2cm} \times \left(\hat{P}^{(\ell')}_{XY,\alpha}(k_j) - \bar{P}_{XY}^{(\ell')}(k_j)\right), 
	\label{eq:cov_def}
\end{align}
where $N_r$ is the number of subvolume realizations, i.e., $N_r=1280$ in our case. Note that we do not include the correction factor in \citet{2007A&A...464..399H}, as it is only 2\% effect in the covariance given a sufficient number of the realizations.
The covariance estimated in this way includes the contribution of the SSC covariance, and therefore serves as an estimator of the total covariance given in Eq.~(\ref{eq:cov_ssc}). 
In the following, we scale the covariance by a factor of $(250~h^{-1}~{\rm Mpc}/1000~h^{-1}~{\rm Mpc})^3$ to approximately obtain the covariance for the volume of $1~(h^{-1}~{\rm Gpc})^3$, a typical volume of ongoing galaxy surveys such as the SDSS BOSS survey\footnote{Exactly speaking, the SSC covariance does not scale with a survey volume as $1/V$, and more rapidly decreases than the scaling. However, the relative decrease compared to $1/V$ is not a strong function of $V$ (a very slowly-varying function of $V$) as shown in Fig.~1 of \citet{2013PhRvD..87l3504T}. The $S/N$ value for the case including the SSC contribution might be changed by 5--10\%, but  the discussion here is qualitatively valid.}.
Due to violation of the periodic boundary conditions in the subvolume, the estimated power spectrum is biased by the window function in low $k$ bins. 
We corrected for this bias by multiplying the estimated power spectrum by a factor of $P(k_i)/P_{\rm sub}(k_i)$ in each $k$ bin, where $P(k_i)$ is the power spectrum estimated from the original simulations with periodic boundary conditions \citep[see around Eq.~53 in][for the details]{2014PhRvD..89h3519L}. 

Fig.~\ref{fig:snr_all} shows the cumulative $S/N$ for the halo power spectrum ($P_{\rm hh}$), the monopole and quadrupole moments of cross-power spectrum of halo and $E$ mode ($P_{{\rm h} E}$), and the monopole of $E$-mode auto spectrum ($P_{EE}$) as a function of the maximum wavenumber $k_{\rm max}$. 
We show the results at $z=0.484$ and $0$ in the left and right panels, respectively, and here we consider the halo sample with $M_{\rm h}=[10^{12},10^{12.5}]~h^{-1}M_\odot$. 
First, the $S/N$ values for all the spectra are saturated at $k\gtrsim 0.4~h~{\rm Mpc}^{-1}$ because the shot noise or shape noise is dominated in the covariance. 
Second, the $S/N$ value for the monopole moment of $P_{{\rm h}E}$ can be greater than 200 at $k_{\rm max}\gtrsim 0.3~h~{\rm Mpc}^{-1}$ for a survey volume of $1~(h^{-1}~{\rm Gpc})^3$, and is about 60\% of that for the density power spectrum for the same halo sample, $P_{\rm hh}$.
This is not so bad, and this results imply that we can measure $P_{{\rm h}E}$ from the same galaxy survey in addition to $P_{{\rm hh}}$. 
If the galaxy shapes have a misalignment with the halo shape, the $S/N$ for the galaxy IA spectrum would be smaller than shown in this plot. 
Comparing the left and right panels manifest that the $S/N$ values are higher for higher redshifts, for a halo sample with a fixed mass threshold. 

How important are the connected non-Gaussian covariance and the super-sample covariance important for the results in Fig.~\ref{fig:snr_all}? 
In the following we address this question. First, we can analytically estimate the Gaussian covariance (${\bf C}^{\rm G}$) and then estimate the cumulative $S/N$ for the Gaussian case, which gives a maximum information content of the $S/N$ value we could extract from the observed cosmological field.  
Once the power spectra of ``$X$'' and ``$Y$'' fields ($X,Y=\delta$, {\rm h} or $E$) are given, the Gaussian covariance matrix is given, as shown in \citet{2010PhRvD..81b3503G} \citep[also see][]{2020PhRvD.101b3510K}, by
\begin{align}
	&{\bf C}^{\rm G}_{(\ell\ell')ij} \equiv \frac{\delta_{ij}}{N_{\rm mode}(k_i)}(2\ell+1)(2\ell'+1)
	\int_{-1}^1 \tkrv{\frac{d\mu}{2}} \mathcal{L}_\ell(\mu)\mathcal{L}_{\ell'}(\mu)\nonumber\\
	&\hspace{2cm} \times \left[ \bar{P}_{XY}^2(k_i,\mu) + \bar{P}_{XX}(k_i,\mu) \bar{P}_{YY}(k_i,\mu) \right],~
	\label{eq:cov_gauss_def}
\end{align}
where $N_{\rm mode}(k_i)$ is the number of Fourier modes that are used for the power spectrum estimation at the $i$-th $k$ bin with width $\Delta k$. 
For a mode satisfying $k_i \gg 2\pi/L$, $N_{\rm mode}(k_i) \simeq 4\pi k_i^2\Delta k / (2\pi/L)^3$, where $L$ is the size of survey volume (the side length of simulation box in our case). 
The Gaussian covariance matrix is diagonal, meaning no correlation between different $k$ bins.
Also note that the auto-power spectra of $P_{XX}$ and $P_{YY}$ include the shot noise or the shape noise contribution. 

Furthermore, to study the impact of the connected non-Gaussian covariance (${\bf C}^{\rm cNG}$), we use a different set of simulations; we run a set of 1000 small-box simulations of $250~h^{-1}~{\rm Mpc}$ size, where we employ $512^3$ particles to keep the same particle/force resolution as in the fiducial simulations, but employ the periodic boundary conditions. 
Then, we measure the power spectrum from each small-box realization, and then estimate the covariance matrix similarly to Eq.~(\ref{eq:cov_def}). 
The covariance matrix estimated from the small-box simulations does not include the SSC contribution, but does includes the contributions of ${\bf C}^{\rm G}$ and ${\bf C}^{\rm cNG}$ in Eq.~(\ref{eq:cov_ssc}).

Fig.~\ref{fig:snr_ssc} shows the $S/N$ values of $P_{{\rm h}E}$ obtained by using the full covariance matrix, the Gaussian covariance matrix (${\bf C}^{\rm G}$) alone, and the covariance matrix without the super-sample covariance contribution (${\bf C}^{\rm G}+{\bf C}^{\rm cNG}$), in the calculation of Eq.~(\ref{eq:snr_ssc}). 
First, all the results fairly well agree with each other up to $k_{\rm max}\simeq 0.2~h~{\rm Mpc}^{-1}$, meaning that the Gaussian covariance is a good approximation up to this wavenumber. 
Second, comparing the gray and red points tells us the the connected non-Gaussian covariance is significant and reduces the $S/N$ value by about 10, 20 and 30\% at $k_{\rm max}\simeq 0.3$, 0.5 and $0.8~h~{\rm Mpc}^{-1}$, respectively. 
Third, comparing the red and blue points, we can find that the SSC further reduces the cumulative $S/N$ value by up to 20\% at $k_{\rm max}\gtrsim 0.2~h~{\rm Mpc}^{-1}$, meaning that the SSC gives a significant contribution to the total covariance at the nonlinear scales. 
The 20\% loss corresponds to about 40\% smaller survey volume as $S/N$ scales roughly with the volume as $S/N\propto V^{1/2}$.
The relative importance of SSC to other covariance terms looks similar to the case of weak lensing covariance \citep{2009ApJ...701..945S,2009MNRAS.395.2065T,2013PhRvD..87l3504T}.
In other words, the SSC term needs to be taken into account if one properly uses the IA power spectrum for cosmology. 
To further study the SSC effect, the separation simulation technique using anisotropic expansion in the local background would be useful \citep{2020arXiv200306427S,2020arXiv200310052M}.

\subsection{2D vs 3D IA power spectrum}
\label{subsec:2d}
\begin{figure}
	\includegraphics[width=0.95\columnwidth]{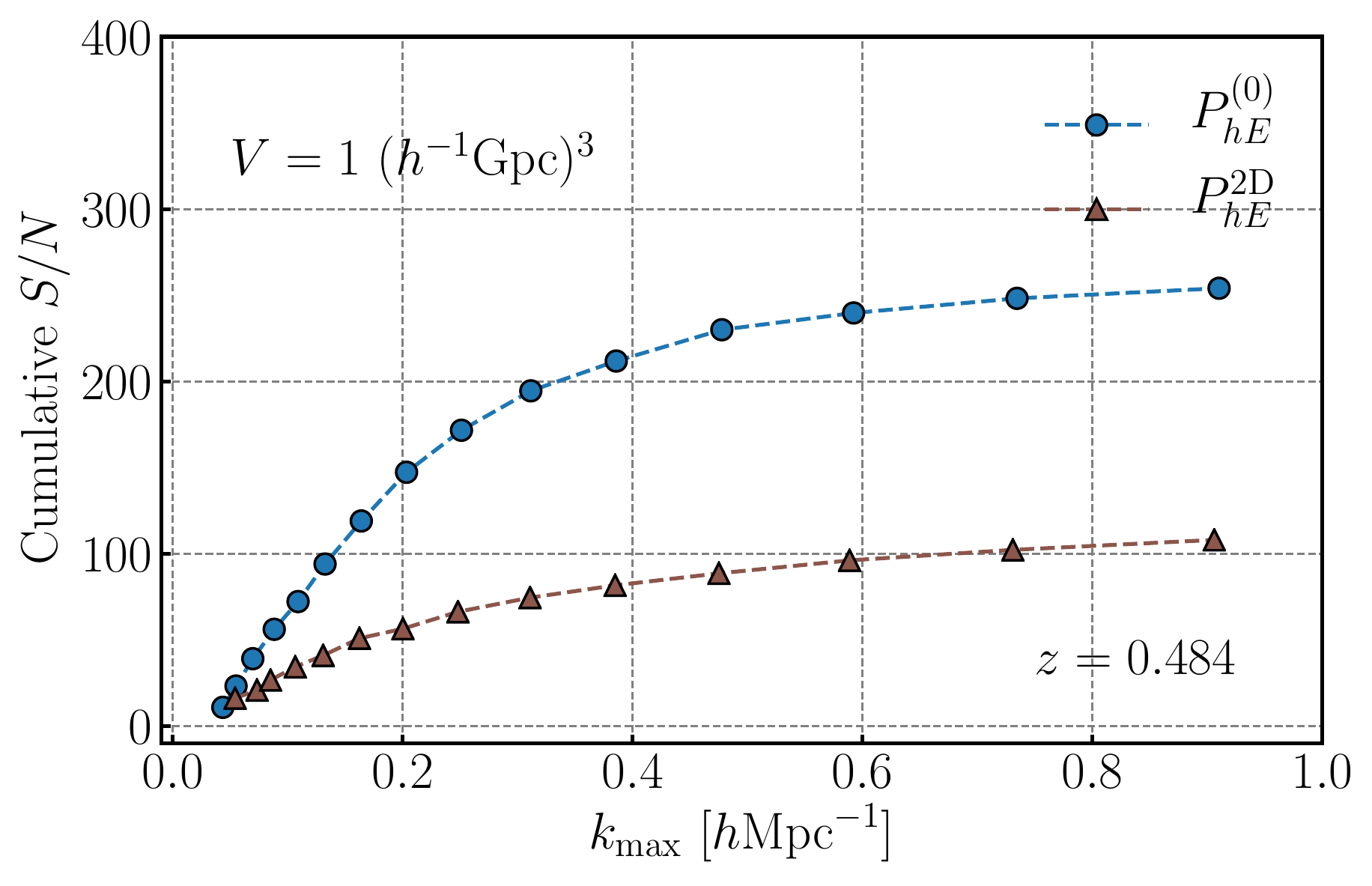}
	\caption{The cumulative $S/N$ for the cross-power spectrum for the projected fields of halos and $E$ mode, $P^{\rm 2D}_{{\rm h}E}$, compared to that for the monopole moment of the 3D power spectrum $P^{(0)}_{{\rm h}E}$ in Fig.~\ref{fig:snr_all}. 
	To define the projected fields, we consider the redshift slice centered at $z=0.484$ and with radial width of $250~h^{-1}~{\rm Mpc}$. 
	To have a fair comparison between the 2D and 3D case, we assume a survey volume of $1~(h^{-1}~{\rm Gpc})^3$ for both cases, where the geometry of 2D case corresponds to $(2000)^2\times (250)~({h}^{-1}{\rm Mpc})^3$. We use the 1280 subboxes to compute the 
	covariance matrix for the 2D spectrum, 
	and the covariance includes the full contributions including the SSC covariance (see text for details).}
	\label{fig:snr_prj}
\end{figure}
We have so far assumed that both three-dimensional positions and shapes of halos are available. 
This is the case that \mtrv{the IA power spectrum measurement is done from }
imaging and spectroscopic galaxy surveys \mtrv{that cover}
the same patch of the sky.
With the advent of deep wide-area multi-band imaging surveys such as the Subaru HSC survey \citep{2018PASJ...70S...4A}, the 
Kilo-Degree survey \citep[KiDS;][]{2015MNRAS.454.3500K}, the Dark Energy Survey \citep[DES;][]{2018PhRvD..98d3526A,2016PhRvD..94b2002B}, the Rubin Observatory's Legacy Survey of Space and Time \citep[LSST;][]{2009arXiv0912.0201L}, Euclid \citep{2011arXiv1110.3193L} and WFIRST \citep{2015arXiv150303757S}, it is natural to ask whether photometric surveys can be used for the IA power spectrum measurements, where the precise radial position (or distance) of individual halos (galaxies) is not available.
To address this question, in this section we investigate how uncertainties in the galaxy redshifts affect our results. 
Here we define the projected shear field as
\begin{align}
    \gamma^{\rm 2D}_{ij}(\bm{x}_\perp) \equiv \int\!\!\mathrm{d}x_3~p(x_3)\gamma_{ij}(\bm{x}_\perp,x_3),
\end{align}
where $p(x_3)$ is the radial selection function satisfying the normalization condition, $\int_0^\infty\!\!\mathrm{d}x_3~p(x_3)=1$.
We employ a simple radial function given by $p(x_3)=1/\Delta\chi$ for $\bar{\chi}-\Delta\chi/2\le x_3\le \bar{\chi}+\Delta\chi/2$, and otherwise $p(x_3)=0$, where $\bar{\chi}$ is the mean comoving distance to the survey slice (survey volume) and $\Delta\chi$ is the width of the redshift slice. 
We define $E$/$B$ modes similarly to Eqs.~(\ref{eq:gammaE}) and (\ref{eq:gammaB}) because the shear field is defined in the two-dimensional plane perpendicular to the line-of-sight direction. 
The power spectrum of the projected field, e.g., the cross-power spectrum of the projected halo and $E$-mode fields is given by
\begin{align}
    \left\langle E^{\rm 2D}(\bk_\perp)\delta^{\rm 2D}_{\rm h}(\bk_\perp^\prime)\right\rangle\equiv P^{\rm 2D}_{{\rm h}E}(k_\perp)(2\pi)^2\delta_D^2(\bk_\perp+\bk_\perp^\prime),
\end{align}
where $\delta_D^2(\bk)$ is the two-dimensional Dirac function. 
As can be found in \citet{2019MNRAS.482.4253T} (see Eq.~29 in their paper), the 2D power spectrum is related to the monopole moment of the 3D power spectrum as
\begin{align}
    P^{\rm 2D}_{{\rm h}E}(k_\perp)\simeq \frac{1}{\Delta\chi}P_{{\rm h}E}(k=k_\perp; z=\bar{z}).
\end{align}
Here we used the notation ``$\simeq$'' because the above equation is exact if we can ignore time evolutions of the fields within the redshift slice we consider (under the distant observer approximation). 
The prefactor, $1/\Delta\chi$, in the above equation accounts for the fact that the fluctuation fields are diluted after the radial projection. 
Here we consider the projected wavenumber $k_\perp$ for comparison purpose with the 3D power spectrum, and the 2D power spectrum is related to the angular power spectrum if the projected field is defined on the celestial sphere, via $C_{{\rm h}E}(\ell)=(1/\bar{\chi}^2)P^{\rm 2D}_{{\rm h}E}(k_\perp=\ell/\bar{\chi})$. 
Hence the following results for the 2D power spectrum are equivalent to what we have for the angular power spectrum.

To have a quantitative comparison of the information contents in the 3D and 2D IA power spectra, we consider the following specifications for a hypothetical imaging survey. 
We consider the mean redshift for $\bar{z}=0.484$, corresponding to $\bar{\chi}=1278~h^{-1}~{\rm Mpc}$ for the {\it Planck} cosmology, and a redshift slice with radial width $\Delta\chi=250~h^{-1}~{\rm Mpc}$ around $\bar{z}$. 
Recalling the relation $\Delta\chi\simeq \Delta z/H(\bar{z})$, the radial width corresponds to the redshift width $\Delta z/(1+z)\simeq 0.074$. 
Although we here consider a top-hat selection around $\bar{\chi}$ for simplicity, the radial selection roughly corresponds to a photo-$z$ accuracy of $\sigma_z\sim 0.04$ on individual galaxies, if we assume that the radial selection corresponds to the $2\sigma$ width of photo-$z$ errors. 
This is comparable to or slightly better than the typical photo-$z$ accuracy for red galaxies as found in the ongoing imaging surveys such as the Subaru HSC survey \citep{2018PASJ...70S...9T}. 
As we did for Fig.~\ref{fig:snr_all}, we divide each simulation of $1~(h^{-1}~{\rm Gpc})^3$ into 64 subboxes each of which has a size of 250~$h^{-1}~{\rm Mpc}$ on a side. 
Then we first project the halo and shear fields along the $x^3$-axis to define the projected fields, and compute the 2D power spectrum from each subbox. 
We then compute the covariance from the 1280 suboxes. 
To have a fair comparison, we scale the covariance to that for a volume of $1~(h^{-1}~{\rm Gpc})^3$, corresponding to a geometry of $1~(h^{-1}~{\rm Gpc})^3=(2000)^2\times (250)~(h^{-1}~{\rm Mpc})^3$, where 250$~h^{-1}~{\rm Mpc}$ is the radial width. The covariance matrix estimated in this way includes all the contributions including the SSC covariance \citep[see also][]{2019MNRAS.482.4253T}.

In Fig.~\ref{fig:snr_prj} we compare the cumulative $S/N$ values for the 2D and 3D cross power spectra of the halo density field and $E$ mode.
The 2D power spectrum has about only a halved information of the 3D spectrum due to the number of available Fourier modes at a certain $k$-bin in the 2D Fourier space compared to the 3D case. 
Thus a spectroscopic survey is advantageous to explore the IA signals. 
In order to explore the full IA information at the level of two-point statistics, we need both imaging and spectroscopic surveys for the same region of the sky. 
As we describe above, the $S/N$ value for the angular IA power spectrum is the same as that of 2D spectrum in Fig.~\ref{fig:snr_prj}.

\subsection{Dependences of the IA power spectra on cosmological parameters}
\label{subsec:cosmology}
\begin{figure*}
	\includegraphics[width=1.9\columnwidth]{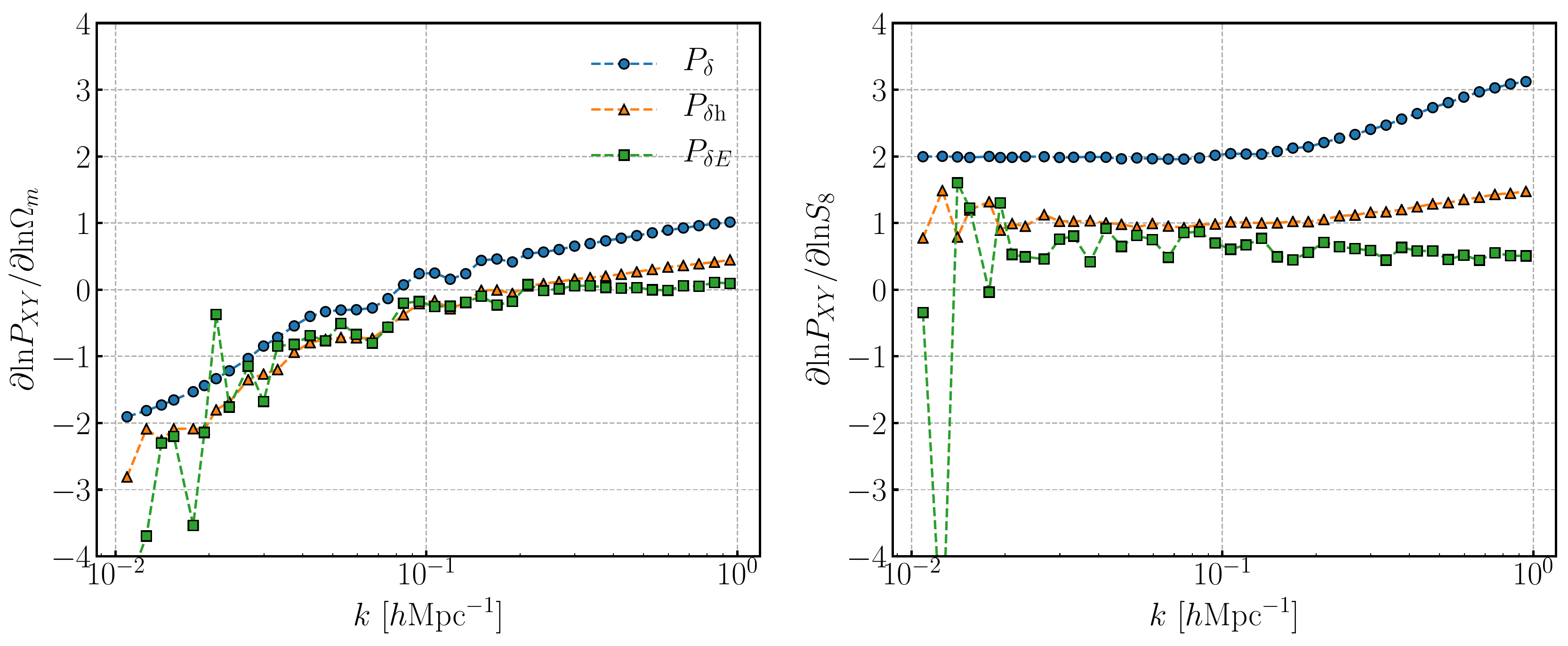}
	\caption{
	\tkrv{
	The dependence of the IA power spectra, $P_{\delta E}$, on the cosmological parameters $\Omega_{\rm m}$ and $S_8$ where $S_8 \equiv \sigma_8(\Omega_{\rm m}/0.3)^{0.5}$.
	Here we consider the number density-threshold sample with $\bar{n}_{\rm h} = 10^{-3}~(h{\rm Mpc}^{-1})^3$ at $z=0.484$. 
	For comparison, we also show the dependences for $P_{\delta}$ and $P_{\delta {\rm h}}$.
	}}
	\label{fig:cosmoparas}
\end{figure*}
How does the IA power spectrum varies with cosmological parameters? 
To address this question, we study how the IA power spectrum depends on the two cosmological parameters, $S_8$ and $\Omega_{\rm m}$. 
Here $S_8\equiv \sigma_8(\Omega_{\rm m}/0.3)^{{0.5}}$ is a parameter to characterize the clumpiness of the universe today, and is the primary parameter to which weak lensing or cosmic shear cosmology is the most sensitive \citep{2019PASJ...71...43H}. 
Since the IA effect is one of the most important, physical systematic effects in cosmic shear cosmology, we study how the IA power spectra depend on these parameters. 
To do this, we run a set of $N$-body simulations where either of $S_8$ or $\Omega_{\rm m}$ is shifted from their fiducial value of {\it Planck} cosmology by $\pm 5\%$, but other parameters are kept to their fiducial values. 
Note that, when we vary $S_8$ with $\Omega_{\rm m}$ being fixed to its {\it Planck} value, we vary $\sigma_8$ alone by an amount corresponding to $\pm5$\% change in $S_8$.
We also use the same initial seeds for one particular realization of the {\it Planck} cosmology simulations in order to reduce scatters due to the sample variance. 
Then we compute the fractional variations in the IA power spectra, computed as
\begin{equation}
    \frac{\partial{\rm ln}P_{XY}}{\partial{\rm ln}p_{\rm cosmo}} \simeq \frac{P_{XY}\left[(1+\varepsilon)p_{\rm cosmo}\right] - P_{XY}\left[(1-\varepsilon)p_{\rm cosmo}\right]}{2\varepsilon P_{XY}\left[p_{\rm cosmo}\right]},
    \label{eq:frac_dlnP}
\end{equation}
where $p_{\rm cosmo}=S_8$ or $\Omega_{\rm m}$, $\varepsilon=0.05$, and $X, Y$ are either of halo (h) and/or the IA $E$ mode ($E$), respectively. 
The fractional differences quantify scaling relations of the IA power spectrum with the cosmological parameters in the vicinity of the {\it Planck} cosmology in two-dimensional parameter space of $(S_8, \Omega_{\rm m})$, given by
\begin{equation}
    \tkrv{
    P_{\delta E} \propto S_8^p \Omega_{\rm m}^q.
    }
    \label{eq:Pk_cosmo}
\end{equation}

Fig.~\ref{fig:cosmoparas} shows the results. 
Although the fractional changes look noisy at small $k$ bins, the IA power spectra display characteristic scale-dependent responses to these parameters. 
The changes get flattened at larger $k$ bins, meaning that changes in these parameters cause an almost scale-independent change in the IA power spectra, just like an overall factor.
The value of each curve in $y$-axis roughly gives the scaling indices $p$ or $q$ in Eq.~(\ref{eq:Pk_cosmo}) at each scale of $k$ bins. 
For the impact of IA effect on the cosmic shear power spectrum for cosmological models around the {\it Planck} cosmology, one needs to further take into account the dependence of the prefactor in Eq.~(\ref{eq:def_of_AI}), $\Omega_{\rm m}/D(z)$, on $\Omega_{\rm m}$.

\subsection{IA power spectra in redshift space}
\label{subsec:redshift-space}
\begin{figure*}
    \begin{center}
    	\includegraphics[width=1.9\columnwidth]{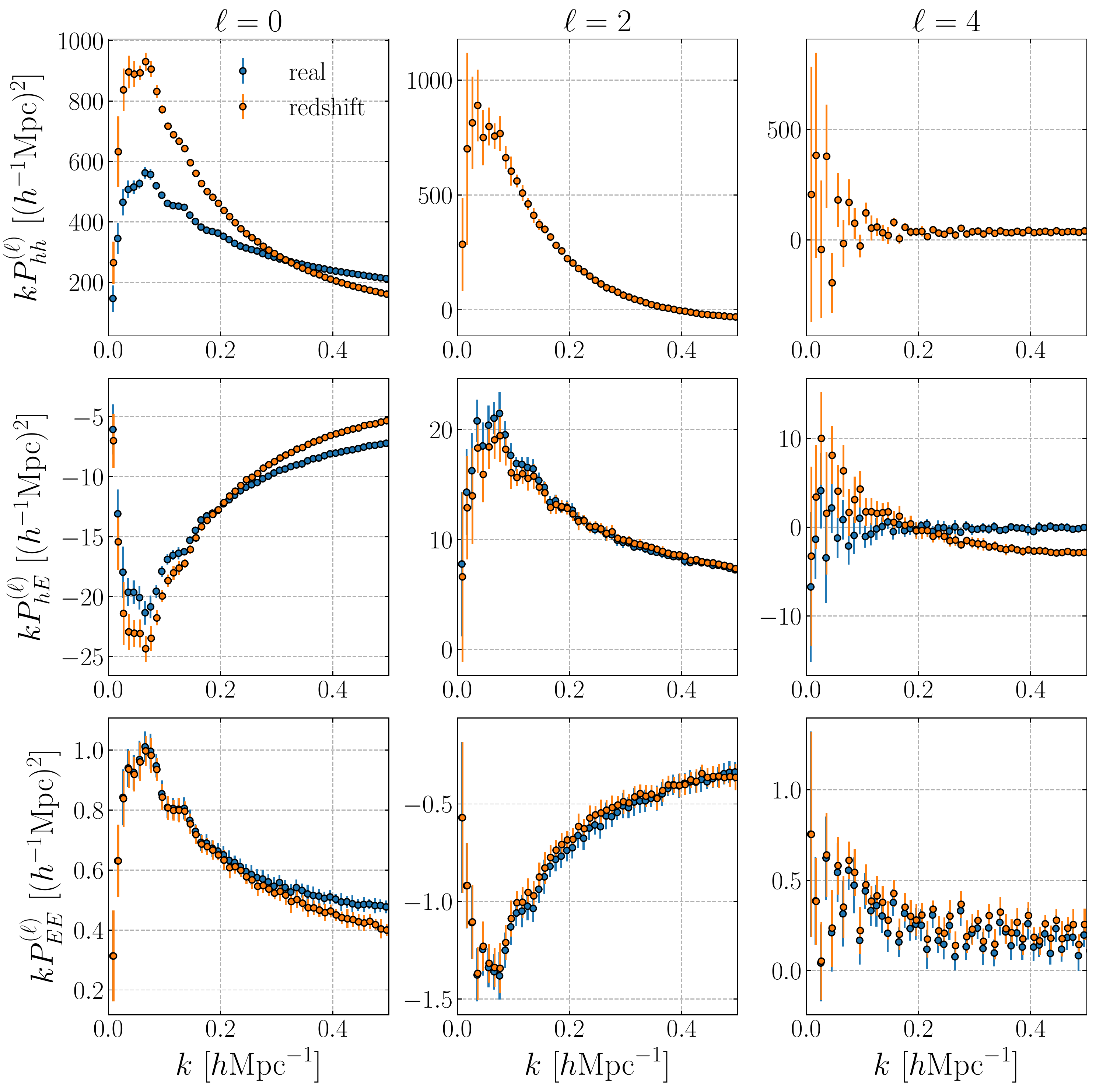}
    	\caption{Comparison between the monopole (left-column panels), quadrupole (middle-column) and hexadecapole (right-column) moments in real and redshift space, for $P_{\rm hh}$, $P_{{\rm h}E}$ and $P_{EE}$.
    	We show the power spectra for the halo sample with $M_{\rm vir} = 10^{12-12.5}~h^{-1}M_{\odot}$ at $z=0.484$.
    	}
    	\label{fig:rsd}
    \end{center}
\end{figure*}
We have so far considered the real- or configuration-space fields. 
However, actual observables for a spectroscopic survey are not real-space fields, but rather defined in redshift space. 
Redshift-space distortions (RSD) due to peculiar velocities of galaxies (halos in our case) \citep{Kaiser1987} cause the observed positions of halos to be modulated compared to those in real space. 

Compared to the standard RSD effect on halos' positions, halo shapes are not affected by the RSD effect \citep{Singhetal2015,2020MNRAS.493L.124O}. 
That is, the shear field in redshift space is invariant under a mapping between real and redshift space:
\begin{equation}
    \boldsymbol{\gamma}^{S}\!(\bs)=\boldsymbol{\gamma}^{R}\!(\bx),
    \label{eq:gamma_S}
\end{equation}
where quantities with superscripts ``$S$'' and ``$R$'' denote the quantities in redshift and real space, respectively, the real- and redshift-space mapping is given by $s_1 = x_1, s_2 = x_2, s_3 = x_3 + v_3/aH$, and $v_3$ is the line-of-sight component of peculiar velocity. 
As we discussed around Eq.~(\ref{eq:gamma_weighted}), however, the shear field estimated from a survey is sampled only at halo's positions, and is affected by the RSD effect on the density field of halos as
\begin{equation}
    \hat{\boldsymbol{\gamma}}^{S}\!(\bs) = \left[1+\delta_{\rm h}^{S}\!(\bs)\right]\boldsymbol{\gamma}^{S}\!(\bs).
\end{equation}
On large scales in the linear regime, the redshift-space density fluctuation field of halos is expressed as
\begin{align}
\delta^S_{\rm h}(\bk)=(1+\beta\mu^2)\delta^R_{{\rm h}}(\bk), 
\label{eq:delta_S}
\end{align}
where $\beta$ is the RSD distortion parameter, defined as $\beta\equiv (1/b)\mathrm{d}\ln D/\mathrm{d}\ln a$. 
The multiplicative factor $1+\beta\mu^2>0$ leads to a boost in the amplitude of redshift-space density fluctuation field compared to the real-space density field on large scales (small $k$). 
Eq.~(\ref{eq:gamma_S}) tells that the RSD effect on the shear field arises from the nonlinear term of fluctuation fields, $\delta^S\gamma^S$. 
Hence the observed shear field on large scales in the linear regime, where $|\delta_{{\rm h}}|\ll 1$, is equivalent to the real-space shear field, i.e., free of the RSD effect. 
However, on smaller scales the observed shear field is affected by the RSD effect, and receives additional $\mu$-modulations, giving characteristic anisotropic patterns in the observed IA shear field \citep[see][for the study on the IA correlation functions in configuration space]{Singhetal2015,2020MNRAS.493L.124O}.

In Fig.~\ref{fig:rsd}, we study the multipole moments of IA power spectra in redshift space, compared to the real-space IA spectra. 
To compute the RSD effect on the halo distribution in simulations, we adopt the bulk motion of each halo that is estimated from the average of velocities of $N$-body particles in a core region of each halo \citep[see][for details]{2020PhRvD.101b3510K}.
As we described, the monopole moment of the redshift-space auto-power spectrum of $E$ mode, $P^{(0)}_{EE}$, is the same as that of the real-space power spectrum on large scales (small $k$) as expected. 
On the other hand, the monopole moment of the redshift-space cross spectrum of halo and $E$-mode fields, $P^{(0)}_{{\rm h}E}$, receives a boost in the amplitude due to the RSD effect, similarly to the effect on the halo power spectrum. 
The RSD effect leads to a non-vanishing hexadecapole moment ($\ell=4$) for $P^S_{{\rm h}E}$, and similarly non-vanishing higher-order moments beyond $\ell=4$ for $P^S_{EE}$.
On small scales in the quasi- and deeply-nonlinear regime, the nonlinear RSD effects cause additional scale dependence in the redshift-space power spectra.

\section{Discussion and Conclusions}
\label{sec:conclusion}

In this work we have developed a novel method to measure the three-dimensional IA power spectra from shapes of halos (as a proxy of galaxy shapes) using a suite of high-resolution $N$-body simulations for the {\it Planck} cosmology. Our findings are summarized as follows:
\begin{itemize}
\item The Fourier-space analysis of halo shapes allows for a straightforward decomposition of the halo shapes into the $E$- and $B$- modes, as in the CMB polarization field and the cosmic shear field.

\item The IA power spectra (the cross spectra of the halo density field and the IA $E$ mode and the auto spectrum of the $E$ mode) display non-vanishing amplitudes on all scales from the linear to nonlinear regimes. 
This means that the primordial fluctuations and gravity in large-scale structure induce a correlation between halo shapes and the matter distribution and between the shapes of different halos on scales much greater than a size of halos (scales of physics inherent in halo formation, a few Mpc at most). 
The IA power spectra on large scales are related to the matter power spectrum, with a \mtrv{scale-independent}
coefficient, as in the linear bias relation of the halo distribution relative to the matter distribution (Figs.~\ref{fig:Pds} and \ref{fig:Pss}). 
This IA constant coefficient ($A_{\rm IA}$) is as expected for the tidal (linear) alignment model for the adiabatic initial condition in $\Lambda$CDM model which we employ for the $N$-body simulations.
The IA shear amplitude is about a few percent at $k\sim 0.1~h~{\rm Mpc}^{-1}$, compared to the intrinsic halo shape of $\gamma_{\rm int}\sim 0.2$ (Fig.~\ref{fig:e_pdf}).
Hence the IA power spectrum can be used to probe the underlying matter power spectrum, very much like what is done using the 
power spectrum \mtrv{of galaxy or halo number density field.}

\item The negative sign of the cross power spectrum of halo density and $E$ mode means that the major axis of halo shapes tend to be {\it statistically} aligned with the minor axis of the tidal field, i.e., the direction of mass accretion onto the halos, which is consistent with the previous simulation results.

\item The IA power spectrum for more massive halos have the greater amplitudes (Fig.~\ref{fig:amp}). 
If we consider the halo sample in a fixed mass bin, the large-scale IA coefficient ($A_{\rm IA}$) asymptotically approaches to a constant value at higher redshift. 
This is as expected for the primordial tidal alignment model \citep{HirataSeljak2004}, implying that the halos shapes of a fixed mass scale at higher redshift retain the information on the primordial tidal field. 
At lower redshifts, the $A_{\rm IA}$ amplitude decreases, probably reflecting the fact that the halo shapes lose the initial memory to some extent due to the mergers or mass accretion in the nonlinear regime. 

\item 
The IA power spectra display BAO features as in the density power spectrum, confirming the similar finding for the real-space IA correlation function \citep{2019PhRvD.100j3507O}. 
In addition, the cross power spectrum of halo density and the IA $E$ mode shows a weaker boost in the amplitude at nonlinear scales compared to the halo density power spectrum, due to the spin-2 nature of the IA field. 

\item The cumulative signal-to-noise ratio ($S/N$) for a measurement of the cross power spectrum of halo density and the IA $E$ mode is about 60\% of that of the halo density power spectrum (Fig.~\ref{fig:snr_all}). 
The super-sample covariance arising from the long-wavelength fluctuations comparable to or greater than a size of survey volume gives a significant contribution to the total covariance as in the covariance of cosmic shear power spectrum (Fig.~\ref{fig:snr_ssc}).
The two-dimensional power spectra of the projected IA field, measured from an imaging survey, suffers from about factor of 2 loss in the information content of the 3D IA power spectrum (Fig.~\ref{fig:snr_prj}). 

\item The IA power spectra in redshift space, the direct observables from galaxy surveys, show additional characteristic anisotropic modulations due to the RSD effects on the halo density field \citep[also see][for the similar discussion]{2020MNRAS.493L.124O}.

\end{itemize}

As we have shown, the IA power spectra can be powerful tools to extract the information on the matter power spectrum, properties of the primordial matter (tidal) perturbations and the cosmological parameters \mtrv{\citep[e.g. see][for such an example]{2020arXiv200703670A}.}
Thus it would be interesting to explore how the IA power spectrum improves the power to constrain cosmological parameters, when combined with the standard density power spectrum. 
This offers additional opportunities that can be attained for imaging and spectroscopic surveys if the two surveys observe the same patch of the sky, where the imaging survey is needed to measure shapes of galaxies and the spectroscopic survey is needed to know the three-dimensional spatial position of the galaxies. 
As we showed, having  spectroscopic redshifts leads to a significant boost in the $S/N$ compared to an imaging survey alone.

In particular, the cross-power spectrum of the galaxy density field and galaxy shapes looks very promising. 
As we showed, the IA shear has the similar amplitudes (a few percent in ellipticities) to the cosmic shear, i.e., weak lensing shear due to large-scale structure in the foreground. 
This would not be surprising because both the effects arise from the gravitational field. 
Even if both imaging and spectroscopic surveys are available, the auto-power spectra of galaxy shapes would suffer from the cosmic shear contamination due to foreground large-scale structures; we cannot distinguish the IA effect and the cosmic shear from the measured power spectra. 
On the other hand, this is not the case for the cross spectrum as long as spectroscopic surveys are available, because the IA cross spectra we are interested in are on scales up to a few 100~$h^{-1}~{\rm Mpc}$ at most, arising from pairs of galaxies separated by such scales (one is for shapes and the other is for the positions) in the common large-scale structure, and the cosmic shear on galaxy shapes by other galaxy would be negligible (recall that cosmic shear builds up by large-scale structures over Gpc scales along the line-of-sight direction).
Since galaxy shapes at higher redshifts might retain more information on the primordial tidal fields (higher $A_{\rm IA}$ coefficients), imaging and spectroscopic surveys for higher redshifts might be more powerful tools of cosmology from joint measurements of the galaxy density and IA power spectra in redshift space. 
Such high-redshift galaxy surveys are, for example, the Subaru HSC and PFS surveys \citep{2014PASJ...66R...1T}.
These are all interesting directions, and are our future work.

\section*{Acknowledgements}
We would like to thank Kazuyuki~Akitsu, Elisa~Chisari, Teppei~Okumura, Jingjing~Shi, and Rensei~Tateishi
for useful discussions. 
This work was supported in part by World Premier International
Research Center Initiative (WPI Initiative), MEXT, Japan, and JSPS
KAKENHI Grant Numbers JP15H03654,
JP15H05887, JP15H05893, JP15K21733, JP17H01131, JP17K14273, JP19H00677 and JP20H04723, by Japan Science and Technology Agency (JST) CREST JPMHCR1414, and JST AIP Acceleration Research Grant Number JP20317829, Japan.
TK is supported by JSPS Research Fellowship for Young Scientists and International Graduate Program for Excellence in Earth-Space Science (IGPEES), World-leading Innovative Graduate Study (WINGS) Program, the University of Tokyo.
KO is supported by JSPS Overseas Research Fellowships.
YK is supported by the Advanced Leading Graduate Course for Photon Science at the University of Tokyo.
The $N$-body simulations and subsequent halo-catalog creation for this work were carried out on Cray XC50 at Center for Computational Astrophysics, National Astronomical Observatory of Japan.

\bibliographystyle{mnras}
\bibliography{refs}

\appendix
\section{Density-weighted field}
\label{app:density_weighted_field}

In this section we describe how to make grid assignments of the halo density and shape fields measured in $N$-body simulation realizations. 
Throughout this section, we omit the subscripts $\{+,\times,h\}$ and write $\epsilon_{(+,\times)}(\bm{x}), n_h(\bm{x}), \delta_h(\bm{x})$ as $\epsilon(\bm{x}), n(\bm{x}), \delta(\bm{x})$ for notational simplicity unless specifically mentioned.

First, the halo number density field $\hn(\bm{x})$ can be formally written as
\begin{equation}
    \hn(\bm{x}) = \sum^{N}_{i=1} \delta_D^3(\bm{x} - \bm{x}_i),
    \label{eq:nx}
\end{equation}
where $N$ is the total number of halos and $\bm{x}_i$ is the position of the $i$-th halo. 
Here the mean halo number density is 
\begin{equation}
    \bar{n} = \frac{\displaystyle \int_V {\rm d}^3\bx~ \hn(\bm{x})}{\displaystyle \int_V {\rm d}^3\bx} = \frac{N}{V}.
\end{equation}
By using an arbitrary weighting function $W(\bm{x})$, we can discretize this field, i.e., evaluate it at the grid point,
\begin{equation}
    \bm{x}_{\rm grid} \equiv  \bm{m} L_{\rm grid} ~(\bm{m} \in \mathbb{Z}^3),
\end{equation}
as
\begin{equation}
    \hn(\bm{x}_{\rm grid}) = \int_V {\rm d}^3x~ W(\bm{x}_{\rm grid} - \bm{x}) \hn(\bm{x}),
    \label{eq:nhgrid}
\end{equation}
where $W(\bm{x})$ satisfies the normalization condition $\displaystyle \int_V\! {\rm d}^3\bx W(\bm{x}) = 1$. 
For example, the NGP assignment is given as
\begin{equation}
    W_{\rm NGP}(\bm{x}) = 
    \begin{cases}
        ~1/L^3_{\rm grid} \equiv 1/V_{\rm grid}~&{\rm where}~|x_1|,|x_2|,|x_3| < L_{\rm grid}/2 \\
        ~0~&{\rm otherwise}
    \end{cases},
\end{equation}
and then Eq.~(\ref{eq:nhgrid}) becomes
\begin{equation}
    \hn(\bm{x}_{\rm grid}) = \frac{1}{V_{\rm grid}} \sum_{i \in {\rm grid}} 1 = \frac{N_{h \in \rm grid}}{V_{\rm grid}},
    \label{eq:nhgrid_ngp}
\end{equation}
where $N_{h \in \rm grid}$ is the number of halos in a grid. 
Therefore the halo number density contrast is calculated by
\begin{equation}
	\hat{\delta}(\bm{x}_{\rm grid}) \equiv \frac{\hn(\bm{x}_{\rm grid})-\bar{n}}{\bar{n}}.
\end{equation}

Next we consider the ellipticity field. 
We have a set of ellipticities of dark matter halos $\{\epsilon_i|i=1, \cdots, N\}$ from a simulation realization and we assume that the ellipticity field is sampled at their position, i.e., $\epsilon_i = \epsilon(\bm{x}_i)$. 
Here we define the discretized ellipticity field in analogy with the density field (Eq.~\ref{eq:nhgrid_ngp}) as
\begin{align}\textbf{}
	\hat{\epsilon}(\bm{x}_{\rm grid}) &\equiv \frac{1}{V_{\rm grid}} \sum_{i \in {\rm grid}} \epsilon_i = \frac{1}{V_{\rm grid}} \sum_{i \in {\rm grid}} \epsilon(\bm{x}_i) \label{eq:tilde_e}\\
	&= \int_V {\rm d}^3\bx~ W_{\rm NGP}(\bm{x}_{\rm grid} - \bm{x})  \epsilon(\bm{x}) \sum^{N}_{i = 1} \delta_D^3(\bm{x} - \bm{x}_i)\\
	&= \int_V {\rm d}^3\bx~ W_{\rm NGP}(\bm{x}_{\rm grid} - \bm{x})  \epsilon(\bm{x}) \hn(\bm{x}).
\end{align}
Therefore $\hat{\epsilon}(\bm{x}) = \epsilon(\bm{x}) \hn(\bm{x})$. 
Finally, by redefining $\hat{\epsilon}(\bm{x}) \to \hat{\epsilon}(\bm{x})/\bar{n}$, we obtain $\hat{\epsilon}(\bm{x}) = (1+\hat{\delta}(\bm{x})) \epsilon(\bm{x})$.
Note that we use the cloud-in-cells (CIC) assignment kernel, a higher order scheme than NGP, in the analyses presented in the main text. This can be achieved simply by replacing $W_\mathrm{NGP}$ with $W_\mathrm{CIC}$ in the above expressions.

\section{Shape noise}
\label{app:shape_noise}
\begin{figure*}
	\begin{minipage}{0.49\hsize}
		\begin{center}
			\includegraphics[width=0.95\columnwidth]{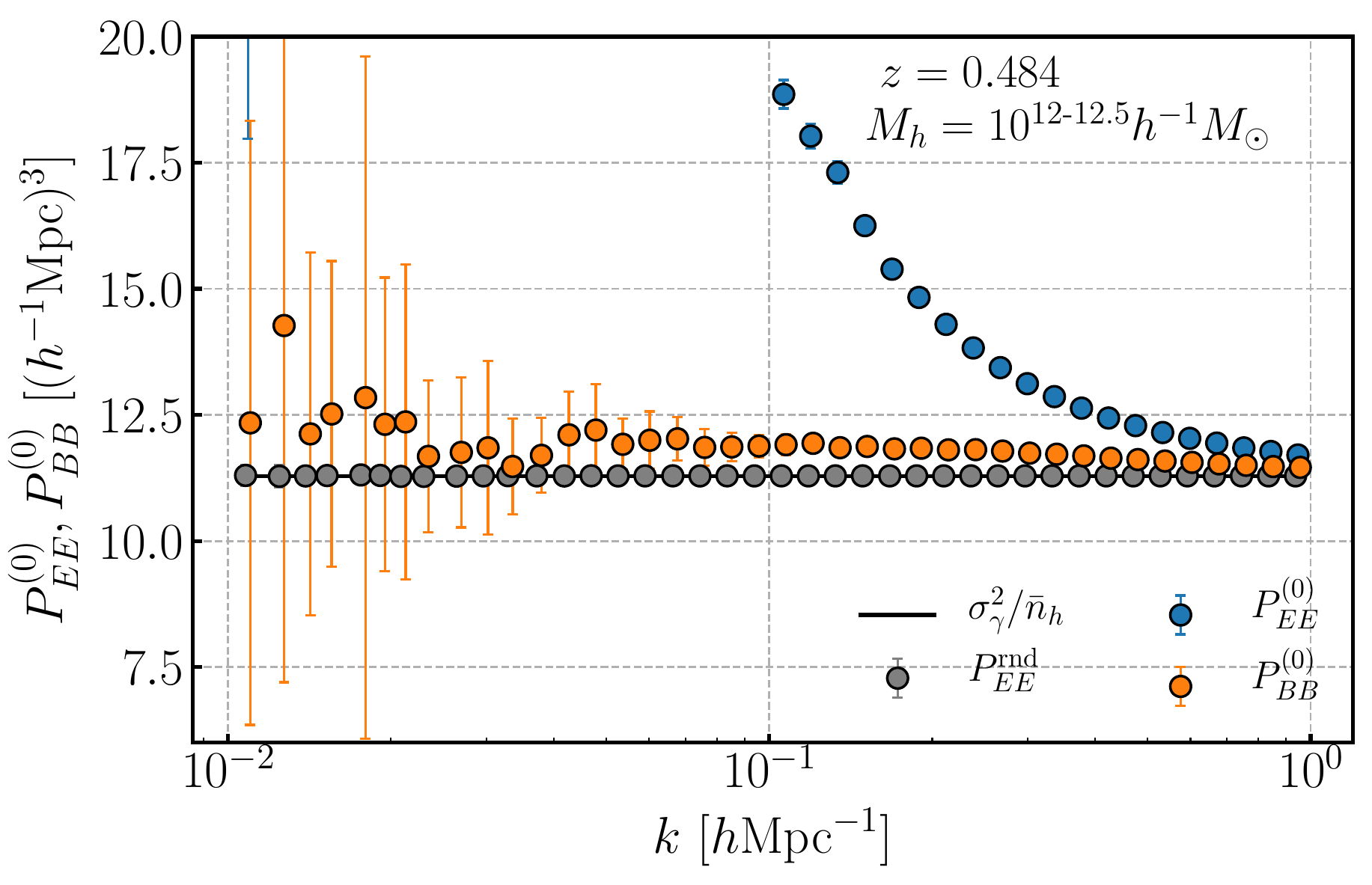}
		\end{center}
	\end{minipage}
	\begin{minipage}{0.49\hsize}
		\begin{center}
			\includegraphics[width=0.95\columnwidth]{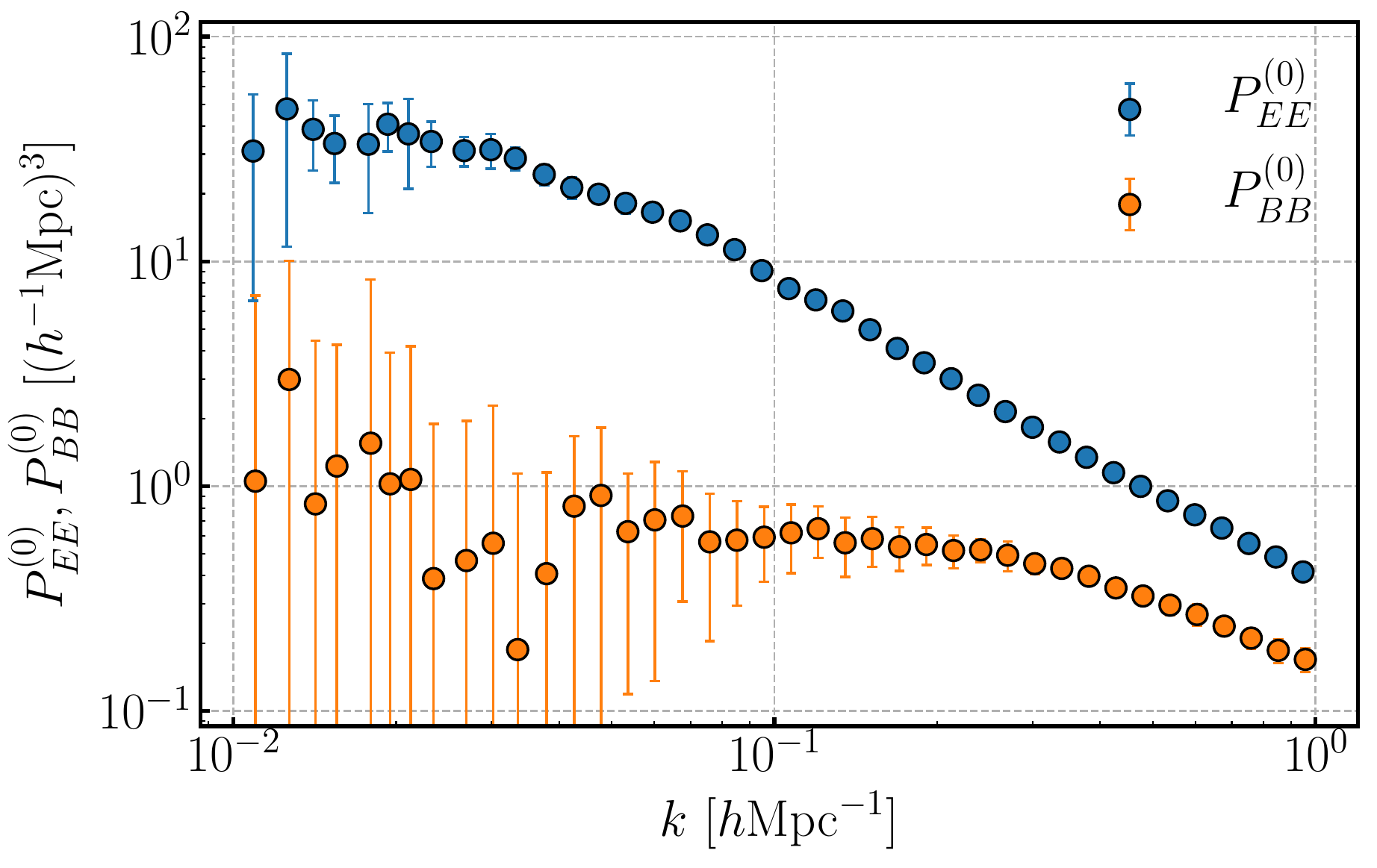}
		\end{center}
	\end{minipage}
	\caption{{\em Left panel}: Comparison between the measured monopole $E$/$B$ auto-power spectra versus the zero lag shape noise. 
	The black line shows a theoretical value of the Poisson shot noise for this halo sample. 
	The gray represents the measured power spectrum after randomly rotating the orientations of the principal axis for all halo samples.
	{\em Right panel}: We show the same auto-power spectra in left panel but after the zero lag subtraction.
 }
 \label{fig:shapenoise}
\end{figure*}
\begin{figure}
	\begin{center}
		\includegraphics[width=0.95\columnwidth]{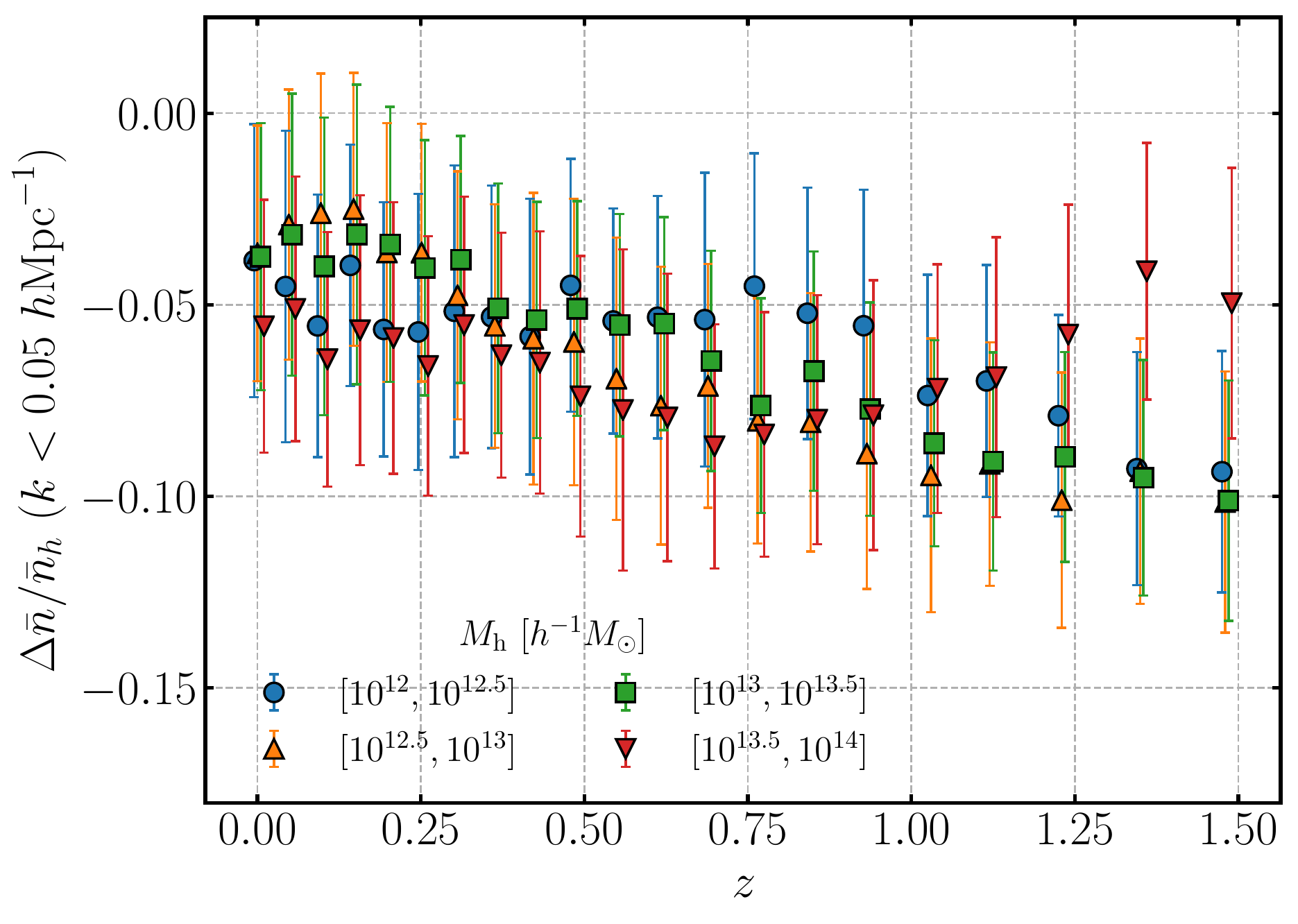}
		\caption{The relative difference of the shape noise $\Delta \bar{n}/\bar{n}_h \equiv \bar{n}_{\rm eff}/\bar{n}_h - 1$. 
		The errorbars represent $1\sigma$ error.
		}
		\label{fig:nef}
	\end{center}
\end{figure}
Here we discuss the shape noise. 
The measured auto-power spectra of halo shape $E,B$ fields, $P_{XX}~(X=\{E,B\})$, have the shape noise contribution that arises due to a finite number sampling of the shape fields at the halo positions.
Unlike the cosmic shear field, there are two contributions.
One is the standard Poisson shot noise term that arises when shapes of different halos are completely uncorrelated, and corresponds to the shape noise term in the cosmic shear power spectrum \citep{2019PASJ...71...43H}. 
The other is from the non-linear evolution of IA \citep{Blazeketal2017}. 
The IA power spectrum itself arises from physical correlation of halo shapes and halo distribution in the same large-scale structure, and this non-Poisson shot noise term contributes the total shot noise term. 
Taking advantage of the spin-2 field of halo shape field, we can disentangle the two contributions. 
This is also the case for actual observations, and is not the case for the density power spectrum. 
One way to estimate the Poisson shot noise is as follows; first, rotate orientation of individual halo ellipticity with random angle, measure the power spectrum in the same way to actual measurements, repeat the random-orientation measurements many times, and then estimate the variance from the many realizations. 
This erases correlated IA effects between different halos keeping the distribution of halos (keeping the clustering of halos). 
In an actual observation, this method can automatically take into account the effects of masks and boundary of survey footprints \citep{2019MNRAS.486...52S}. 
We perform the random-orientation measurements 10000 times for each of 20 each simulation realizations and calculate the mean and variance of the measured power spectra. 
We show the result as $P^{\rm rnd}_{EE}$ by gray points in Fig.~\ref{fig:shapenoise}. 
We does not show $P^{\rm rnd}_{BB}$ because this is almost the same as $P^{\rm rnd}_{EE}$. 
We find that $P^{\rm rnd}_{EE}$ is in good agreement with a theoretical Poisson shot noise $\sigma^2_\gamma/\bar{n}_h$ shown in the black line, where the intrinsic shape rms, $\sigma_\gamma$, is estimated from the distribution of halo ellipticities in Fig.~\ref{fig:e_pdf} taking into account the responsivity ${\cal R}$.
For comparison, we also show $P^{(0)}_{EE}$ and $P^{(0)}_{BB}$ without subtracting the Poisson shot noise term.
Both the power spectra agree with the Poisson shot noise term at sufficiently large $k$ as expected.

Interestingly the $B$-mode power spectrum shows a clear deviation from the Poisson shot noise. 
The extra contribution is considered as the ``renormalized'' term arising from the $k\to 0$ limit of higher-order terms in the $B$-mode power spectrum \citep{2009JCAP...08..020M} \citep[also see 
][]{Blazeketal2017}.
In particular, it converges to a certain $k$-independent constant in $k\to 0$ limit. 
The difference between the constant values at the limits of $k\to \infty$ and $k\to 0$ could be recognized as the difference between the (bare) number density $\bar{n}_h$ and the effective number density $\bar{n}_{\rm eff}$ which is defined by
\begin{equation}
    \bar{n}^{-1}_{\rm eff} \equiv \frac{1}{\sigma^2_\gamma} \lim_{k \to 0} P^{(0)}_{EE/BB}(k).
\end{equation}
In this work we estimate $\bar{n}_{\rm eff}$ for our halo samples from simulation by minimizing the $\chi^2$ statistics:
\begin{equation}
    \chi^2\equiv \sum_{k_i; k_i<0.05~h~{\rm Mpc}^{-1}}
    \frac{[P^{(0)}_{BB}(k_i) - \sigma^2_\gamma/\hat{\bar{n}}_{\rm eff}]^2}{\sigma_{BB_i}^2},
    \label{eq:n_eff_est}
\end{equation}
where $\sigma_{BB_i}^2$ is the variance of $P^{(0)}_{BB}(k_i)$ of 20 simulation realizations. 
We can safely estimate the constant offset by using $k$ modes in the sufficiently linear regime.
Once again, we should note that the discrepancy from the Poisson shot noise can be estimated from actual data, by comparing the Poisson shot noise, estimated by the above method, and the measured $B$-mode power spectrum.

In Fig.~\ref{eq:n_eff_est} we show the relative difference of the number density, $\Delta \bar{n}/\bar{n}_h \equiv \bar{n}_{\rm  eff}/\bar{n}_h - 1$. 
The non-Poisson shot noise compared to the Poisson shot noise is roughly $5 \text{--} 10$\% for all the halo samples we consider.

\tkrv{
\section{\mtrv{A dependence of IA power spectrum on definition of inertia tensor}}
\label{app:definitions_of_inertia_tensor}
In this section, we \mtrv{study}
how our results \mtrv{vary with different definitions of}
the inertia tensor of individual halo shapes. 
\mtrv{To do this}
we consider \mtrv{eight different definitions of the inertia tensor}
in total which \mtrv{have been used}
to define the shape of a simulated halo or galaxy \mtrv{in the literature.} 
Those are basically identical to the inertia tensors summarized in \cite{2012MNRAS.420.3303B}, but here we briefly review the definition and motivation of each inertia tensor.
First, those definitions \mtrv{are categorized}
into two types; the simple (unweighted) inertia tensor, $I^{\rm Sim}_{ij} \propto \sum_p \Delta x^i_p \Delta x^j_p$, and the reduced (weighted) inertia tensor, $I^{\rm Red}_{ij} \propto \sum_p \Delta x^i_p \Delta x^j_p / r^2_p$ where $\Delta \bx_p$ is the position vector 
\mtrv{of each member particle $p$}
from the halo \mtrv{(or galaxy)}
center 
and $r_p = |\Delta \bx_p|$.
Note that since $\Delta x^i_p \Delta x^j_p/r^2_p = \hat{n}^i_p\hat{n}^j_p$ where $\hat{\bn}$ is a unit vector, all particles are projected onto a unit sphere.
This $1/r^2_p$ weight reduces \mtrv{the contribution of outer particles, e.g.}
the effect of massive subclumps in the outskirts, and then we consider that $I^{\rm Red}_{ij}$ approximates the 
shape of a virialized object around the gravitational potential minimum better than $I^{\rm Sim}_{ij}$, \mtrv{which can be considered as a proxy of shape of a central galaxy if it forms at the center of the halo.
For each \mtrv{inertia tensor,}
we can define the shape as a function of 
\mtrv{the boundary}
radius $R$ by using only particles satisfying $r_p < R$ in the summation $\sum_p$ \mtrv{or}
using all member particles defined in the {\tt Rockstar} halo finder.
In this work, we set $R$ as the virial radius, $R = r_{\rm vir}$.
\mtrv{Hereafter}
we \mtrv{refer to the former definition}
as `Sim' or `Red' simply, and the latter as `Sim-AllParts' or `Red-AllParts', respectively.}}

\tkrv{\mtrv{However, we find that the measured shape is quite sensitive to the boundary radius, and the shape varies with changing 
the boundary radius, which is ascribed to the spherical boundary.}
Hence we use the iterative method introduced in \cite{1991ApJ...368..325K} for both `Sim' and `Red' cases to \mtrv{mitigate}
this artificial effect. \mtrv{As an example,}
we below describe the algorithm for the reduced inertia tensor.
}
%
%
\tkrv{First, we \mtrv{estimate the inertia tensor $I^{(0)}_{ij}$ and use it as the initial guess of $I^{\rm Red}_{ij}$.}
}
Second, we diagonalize $I^{(0)}_{ij}$ and obtain the eigenvectors $\bm{e}_a,\bm{e}_b,\bm{e}_c$ corresponding to the three principal axes $a,b,c~(a>b>c)$. 
Then we define the 
new radius for each particle as
\begin{equation}
	r^{(1)}_p \equiv \sqrt{\left(\bx_p \cdot \bm{e}_a \right)^2 + \left( \frac{\bx_p \cdot \bm{e}_b}{s}\right)^2 + \left(\frac{\bx_p \cdot \bm{e}_c}{q}\right)^2},
	\label{eq:rp_ac}
\end{equation}
where $q \equiv c/a,s \equiv b/a$ are the axis ratios.
Third, by using member particles which satisfy $r^{(1)}_p < r_{\rm vir}$, we redefine the inertia tensor replacing $r^{(0)}_p$ with $r^{(1)}_p$:
\begin{equation}
	I^{(1)}_{ij} \equiv \sum_p \frac{\Delta x^i_{ p} \Delta x^j_{ p}}{\left(r^{(1)}_p\right)^2}.
	\label{eq:Iij_1}
\end{equation}
We perform the second and third step calculations iteratively until $q$ and $s$ converge to within 1\% precision and we finally use the converged inertia tensor $I_{ij}$, \mtrv{as an estimate of $I^{\rm Red}_{ij}$ for the halo,  
to define the ellipticities.}
\tkrv{
This iterative algorithm \mtrv{is based on the fixed boundary for}
the major axis of the ellipsoid at all steps, i.e., 
$a=r_{\rm vir}$. 
We call this definition
as `Red-Iter-AC', which is our default definition of halo ellipticities used in the main text.  
There is an alternative choice to get the new radius instead of Eq. (\ref{eq:rp_ac}) \citep{2012JCAP...05..030S}:
\begin{equation}
	\frac{r^{(1)}_p}{(abc)^{1/3}} \equiv \sqrt{\left (\frac{\bx_p \cdot \bm{e}_a}{a} \right)^2 + \left( \frac{\bx_p \cdot \bm{e}_b}{b}\right)^2 + \left(\frac{\bx_p \cdot \bm{e}_c}{c}\right)^2}.
	\label{eq:rp_vc}
\end{equation}
This replacement \mtrv{keeps} 
the volume of the boundary ellipsoid 
\mtrv{fixed to}
constant at all steps, i.e., $V=4\pi r^3_{\rm vir}/3$.
Thus we call this definition
as `Red-Iter-VC'.}
\tkrv{
Note that we can also compute the \mtrv{iterated inertia tensor without $1/r_p^2$ weighting},
and call the ellipticities
as `Sim-Iter-AC' and `Sim-Iter-VC', respectively.
}

\tkrv{
\begin{figure}
	\begin{center}
		\includegraphics[width=0.95\columnwidth]{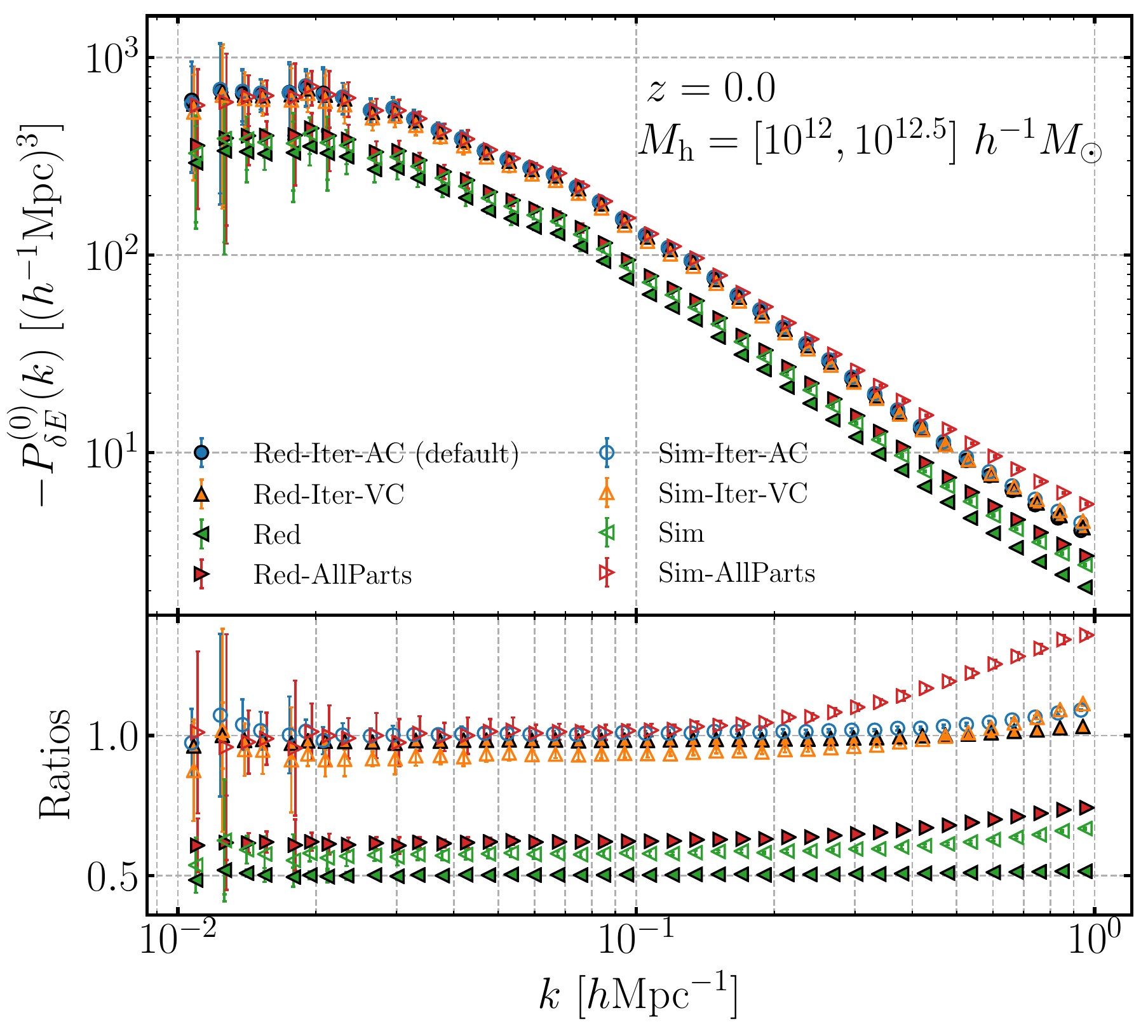}
		\caption{
		{\it Upper panel}: The IA power spectra with various definitions of the inertia tensors.
		{\it Lower}: The ratios of the spectra to the spectrum of `Red-Iter-AC' (our default) case.
		}
		\label{fig:DefsIij_Power}
	\end{center}
\end{figure}
\begin{figure}
	\begin{center}
		\includegraphics[width=0.95\columnwidth]{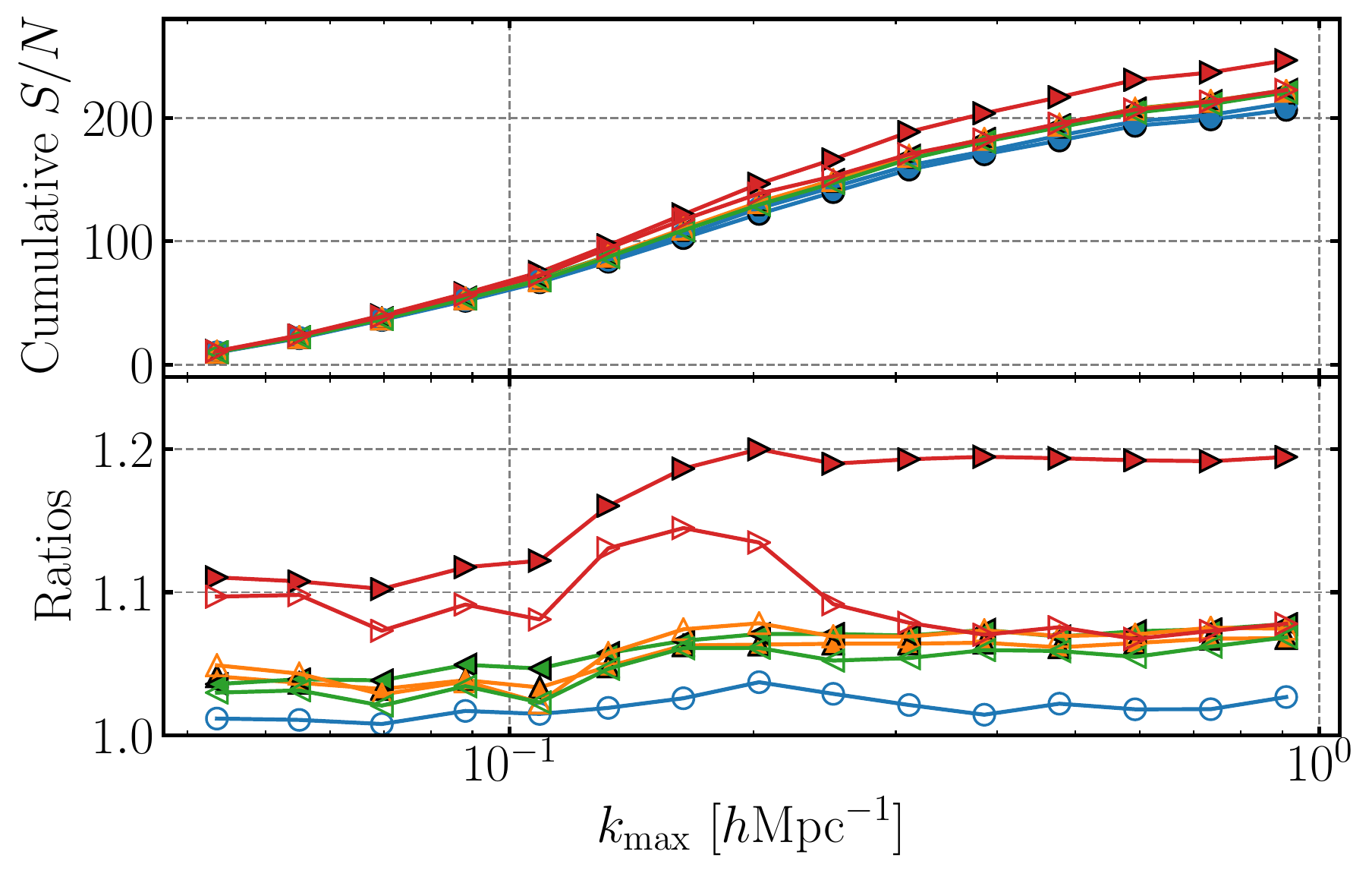}
		\caption{
		{\it Upper panel}: The cumulative $S/N$ of the IA spectra with $V = 1~(h^{-1}{\rm Gpc})^3$.
		{\it Lower}: The ratios of the $S/N$ to the $S/N$ of `Red-Iter-AC' (our default) case.
		}
		\label{fig:DefsIij_SNR}
	\end{center}
\end{figure}
Fig.~\ref{fig:DefsIij_Power} shows the dependences of the shapes ($k$-dependence) of the IA power spectrum on various inertia tensors.
We find that the spectra from different inertia tensors display different constant factors at least $k \lesssim 0.1~h{\rm Mpc}^{-1}$ and also display different $k$-dependence in the nonlinear regime ($k \gtrsim 0.1~h{\rm Mpc}^{-1}$) as a whole.
Comparing the reduced tensors (filled symbols) with the simple tensors (open) for the fixed color (shape of marker), the boost of the simple tensors' case in the nonlinear regime is owing to the particles in the outer region.
In particular, the fact that two `AllParts' signals (red triangle) exhibit relatively higher boosts than the others do is also explained by the mass distribution in the extended region beyond the virial radius.
The spectra from 3 reduced tensors (`Red-Iter-AC', `Red-Iter-VC' and `Red') which includes only inner particles satisfying $r_p < r_{\rm vir}$ are almost the same $k$-dependence even at highly nonlinear scale.
The amplitudes of `Red' and `Sim' (green, left triangle) signals are halved from the iterative signals due to the artificial spherical boundary.
}

\tkrv{
The signal-to-noise ratio ($S/N$) of the spectrum in Fig.~\ref{fig:DefsIij_SNR}, on the other hand, is changed only by 10\% at large scales for different methods
, i.e. $A_{\rm IA}/\sigma_\gamma$ is nearly constant.
In particular, the difference is less than 10\% at all scales for the methods using the particles around the virial radius (blue, orange and green).
Thus, if we measure (define) all halo shapes by using the same scheme self-consistently and if we consider the IA power spectrum 
at larger scales than a typical scale of halos, we obtain almost the same information content from it regardless of the detail of shape measurements.
}

\tkrv{
\section{Dependences on member particle resolutions}
\label{app:particle_resolution}
\begin{figure}
	\begin{center}
		\includegraphics[width=0.95\columnwidth]{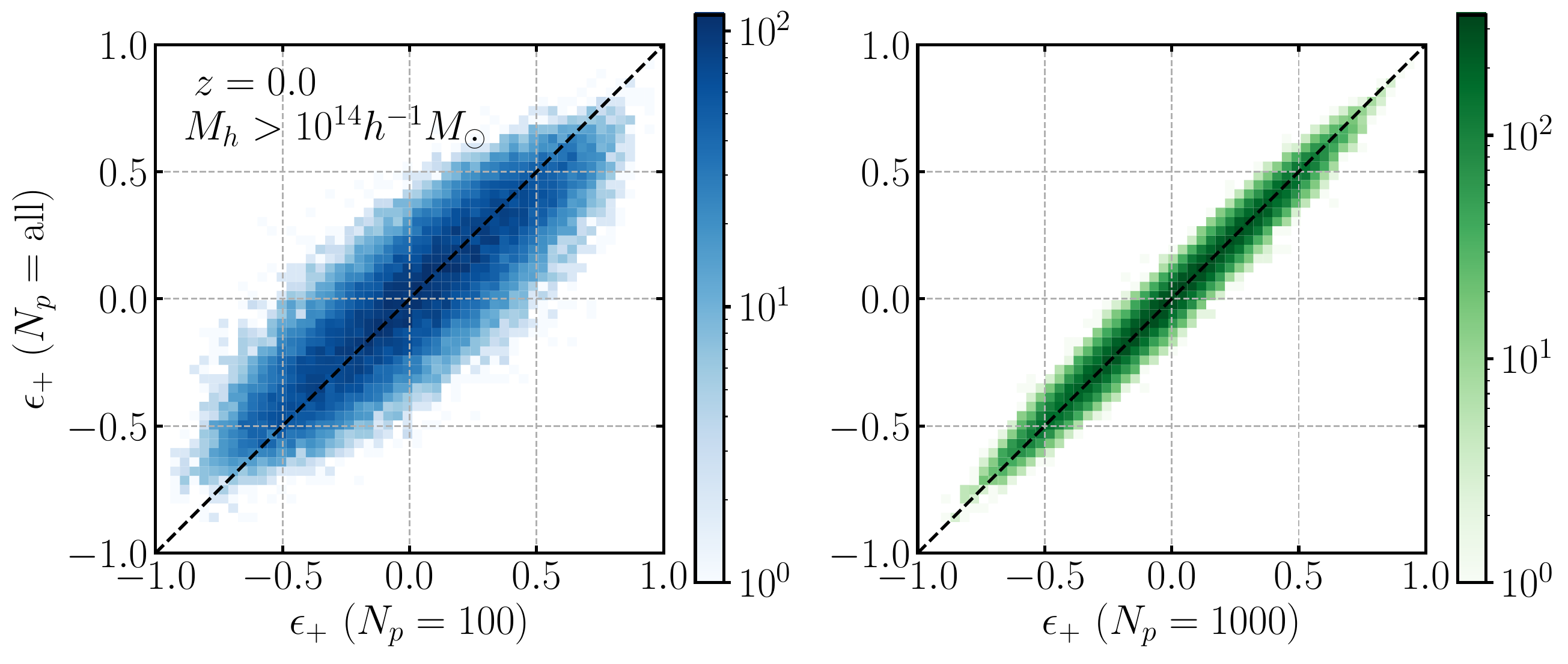}
		\caption{
		The two dimensional histogram between low-resolved halo shapes ($x$-axis) and high-resolved shapes ($y$-axis).
        The left panel is for the $N_p=100$ case and the right is for the $N_p=1000$ case, respectively.
        The color bars indicate the number of halos with a logarithmic scale where $\Delta \epsilon_+ = 0.04$ for the bin width.
        We also show the reference line, $y=x$, in the black dashed line.
		}
		\label{fig:PartRes_Ellip}
	\end{center}
\end{figure}
\begin{figure}
	\begin{center}
		\includegraphics[width=0.95\columnwidth]{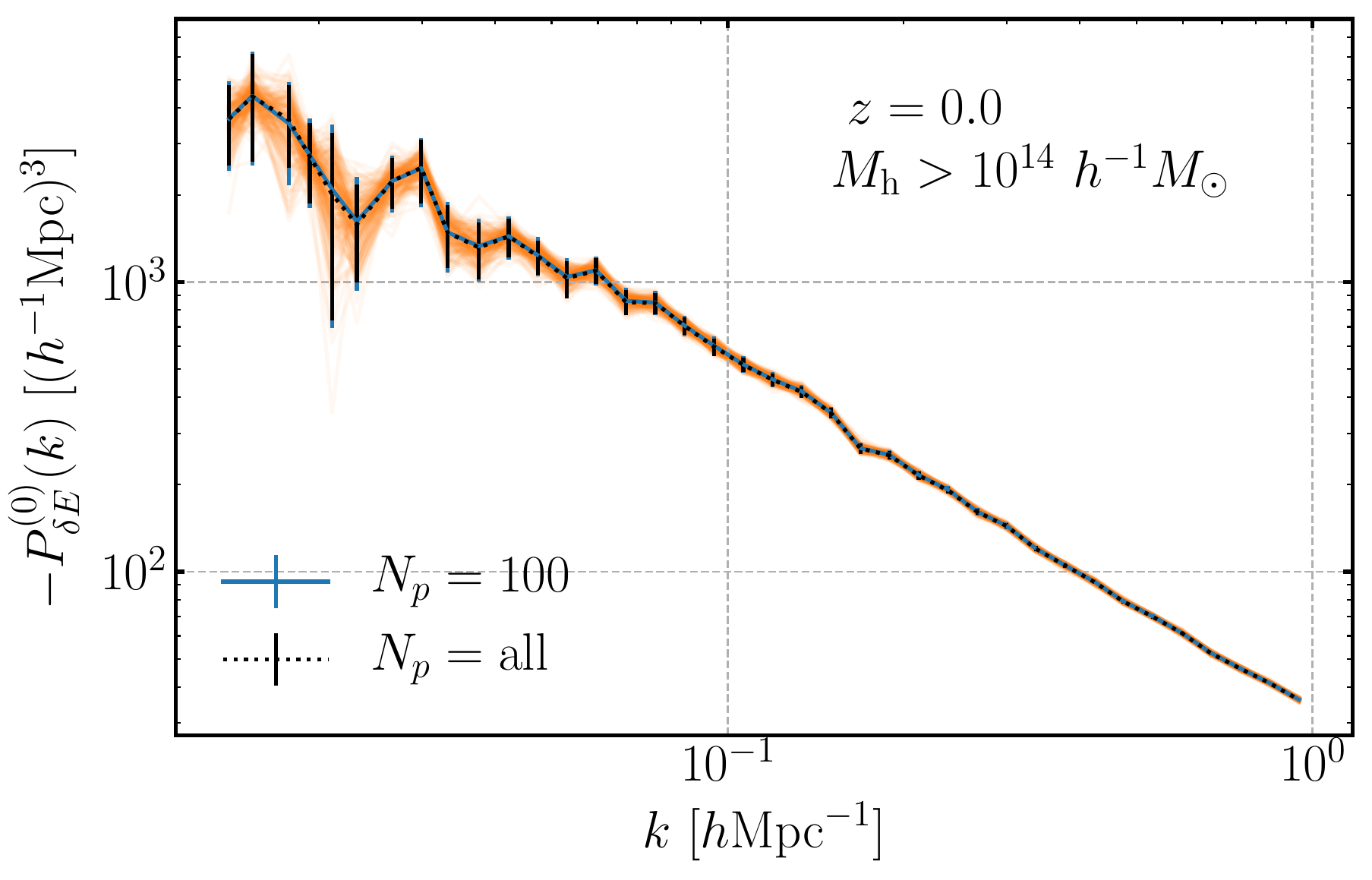}
		\caption{
		The IA power spectra varying the number of member particles to estimate halo shapes for one simulation realization. 
		The black-dotted line shows the original signal, i.e. we used all member particles for $I_{ij}$. 
        The blue-solid line shows the $N_p=100$-signal averaged over 200 times for different 100-particle choices (background orange lines).
        The error bars correspond to the statistical error of the volume $V=1~(h^{-1}{\rm Gpc})^3$.
        The error bars of the $N_p=100$ signal are slightly larger than those of the original signal for each $k$-bin due to the increase of $\epsilon_{\rm rms}$.
		}
		\label{fig:PartRes_Power}
	\end{center}
\end{figure}
As mentioned in Section~\ref{sec:estimate-shear-field}, we use halos with masses down to 
$M_{\rm h} = 10^{12}~h^{-1}M_\odot$, where the minimum halo mass 
roughly corresponds to $100$ member particles in our simulations 
because
the mass of each $N$-body particle is $m_p \sim 1\times 10^{10}h^{-1}M_\odot$.
\mtrv{One might be concerned about an inaccuracy of the inertia tensor definition for such small halos due to a smaller number of member particles.
In this section we study}
how this small number of member particles affects measurements of
the 
IA power spectrum.
In fact, 100 particles are not sufficient to precisely characterize the shapes of {\it individual} halos. 
\mtrv{To study this, we consider very massive halos with $M_h\ge 10^{14}~h^{-1}M_\odot$ that contains more than $10^4$ member particles.}
In Fig.~\ref{fig:PartRes_Ellip} we study how the measurement accuracy of individual halo shapes 
is
degraded if we use a partial fraction of the member particles. The figure shows the scatter plot for the same sample of halos;
the $y$-axis shows the ellipticities when using all the member particles for each halo, 
while the $x$-axis shows the ellipticities for the same halo when using only 100 or 1000 particles randomly selected from the member particles. 
Each plot displays
a huge scatter between the two ellipticity estimates. As a result, 
the root mean square of ellipticities per component among all the halos,  $\sigma_\epsilon$, increases when using a smaller number of the member particles; 
$\sigma_\epsilon = (0.3076, 0.3122, 0.3467)$ for $N_p = ({\rm all}, 1000, 100)$, respectively.
Thus the ellipticity measurements become noisy on individual halo basis. 
The increased random noise of individual halo shapes 
leads to an increase of the statistical shape noise in the IA power spectrum measurements, $\sigma^2_\epsilon / \bar{n}_h$ (about $\sim 20\%$ in the case of $N_p=100$ for instance).
}

\tkrv{
On the other hand, as shown in Fig.~\ref{fig:PartRes_Power}, the measured IA power spectrum is almost unchanged because it carries only the 
physically-correlated shapes between different halos (also see Fig.~\ref{fig:EmodeAmp} for the similar discussion).  
Therefore we conclude that we can safely measure the IA power spectrum even if the shape measurements of individual halos are affected by the random noise, e.g. 
due to the use of a finite number of particles, 
and then an estimate of
the large-scale amplitude $A_{\rm IA}$ is unbiased or not affected by the shape measurement noise, as long as the shapes of halos are defined by the 
same method self-consistently for different halos.
However its fitting error would be slightly 
overestimated, and 
the $S/N$ of the power spectrum would be slightly underestimated in the shape noise dominated regime. 
}

\tkrv{
\section{large-scale amplitude as a nuisance parameter}
\label{app:b_K}
\begin{figure}
	\begin{center}
		\includegraphics[width=0.95\columnwidth]{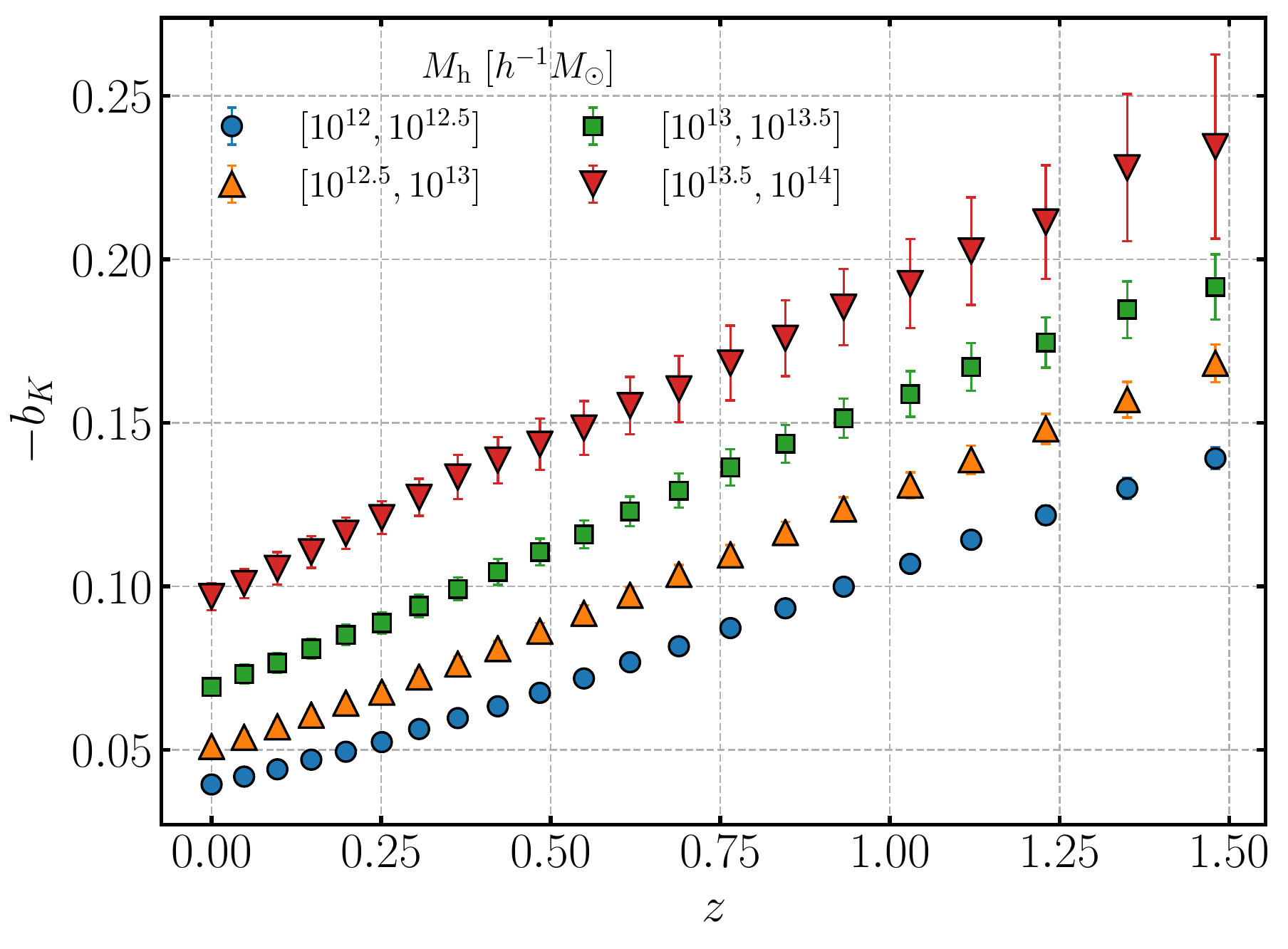}
		\caption{
		The halo mass and redshift dependence of $b_K$.
		This figure is equivalent to the left panel of Fig.~\ref{fig:amp}.
		}
		\label{fig:b_K}
	\end{center}
\end{figure}
In the main text, we adopt $A_{\rm IA}$ which is defined by Eq.~(\ref{eq:def_of_AI}) as the large-scale amplitude of the IA signal to make it easier to compare our results with many previous studies and to discuss its physical origin along
with the {\it primordial alignment} scenario.
However, this is not a unique way to characterize the IA amplitude. 
As claimed in \citet{Schmidtetal2015}, one could define the IA shear field in terms of the tidal field with a linear coefficient 
at large scales: 
\begin{equation}
    g_{ij}(\bx;z) = b_K K_{ij}(\bx;z).
    \label{eq:def_of_bK}
\end{equation}
Here $K_{ij}$ is defined by Eq.~(\ref{eq:Kij_def}) that 
has the same dimension as that of the mass density fluctuation,
%
%
and $g_{ij}$ is the three-dimensional halo shape tensor.
The linear coefficient, $b_K$, in the above equation is dimension-less and 
defined in analogy with the linear density bias parameters, $\delta_{\rm g}=b_1\delta$. 
The coefficient $b_K$ is related to  
$A_{\rm IA}$, which we mainly consider in this paper,  
as $b_K = -A_{\rm IA}C_1\rho_{\rm cr0}\Omega_{\rm m}/D(z)$.
}

\tkrv{For comprehensiveness of our discussion, 
we show the mass and redshift dependences of $b_K$ in Fig.~\ref{fig:b_K}.
Each value is estimated by the same procedure as that for $A_{\rm IA}$ described in Section~\ref{subsec:results_halomassetc}.
}

\bsp	
\label{lastpage}
\end{document}